\documentclass[12pt]{article}
\usepackage[margin=1.25in]{geometry}

\usepackage{natbib}
\usepackage{amsmath, bm,bbm}
\usepackage{amssymb}
\usepackage{graphicx}
\usepackage{multicol}
\usepackage{float}
\usepackage{hyperref}
\usepackage{color}
\usepackage{rotating}
\usepackage{subfig}
\usepackage{enumitem}
\usepackage{mathtools}
\usepackage{setspace}
\usepackage{verbatim}
\usepackage{soul}
\RequirePackage{etoolbox}
\RequirePackage{xcolor}
\definecolor{cb-blue}{RGB}{0, 109, 219}
\definecolor{cb-rose}{RGB}{255, 109, 182}

\DeclarePairedDelimiter\abs{\lvert}{\rvert}%
\makeatletter
\let\oldabs\abs
\def\abs{\@ifstar{\oldabs}{\oldabs*}}

\makeatletter
\def\namedlabel#1#2{\begingroup
    #2%
    \def\@currentlabel{#2}%
    \phantomsection\label{#1}\endgroup
}
\makeatother

\newtheorem{definition}{Definition}

\begin{document} 

\title{\large \bf Private Tabular Survey Data Products through Synthetic Microdata Generation}

\author{Jingchen Hu\footnote{Vassar College, Box 27, 124 Raymond Ave, Poughkeepsie, NY 12604, jihu@vassar.edu} $\,$ and Terrance D. Savitsky\footnote{U.S. Bureau of Labor Statistics, Office of Survey Methods Research, Suite 5930, 2 Massachusetts Ave NE Washington, DC 20212, Savitsky.Terrance@bls.gov} $\,$ and Matthew R. Williams\footnote{RTI International, 3040 East Cornwallis Road, Research Triangle Park, NC 27709, mrwilliams@rti.org}}

\maketitle

\abstract{We propose two synthetic microdata approaches to generate private tabular survey data products for public release. We adapt a pseudo posterior mechanism that downweights by-record likelihood contributions with weights $\in [0,1]$ based on their identification disclosure risks to producing tabular products for survey data.  Our method applied to an observed survey database achieves an asymptotic global probabilistic differential privacy guarantee.  Our two approaches synthesize the observed sample distribution of the outcome and survey weights, jointly, such that both quantities together possess a privacy guarantee. The privacy-protected outcome and survey weights are used to construct tabular cell estimates (where the cell inclusion indicators are treated as known and public) and associated standard errors to correct for survey sampling bias. 
Through a real data application to the Survey of Doctorate Recipients public use file and simulation studies motivated by the application, we demonstrate that our two microdata synthesis approaches to construct tabular products provide superior utility preservation as compared to the additive-noise approach of the Laplace Mechanism. Moreover, our approaches allow the release of microdata to the public, enabling additional analyses at no extra privacy cost.}


\section*{Statement of Significance}

Our work lies in the intersection of differential privacy, synthetic data, and survey data. We focus on providing privacy protection for tabular survey data products using synthetic microdata under differential privacy. Our work proposes two formally private data synthesizers able to account for sampling weights, that outperform the additive noise method of Laplace Mechanism. We demonstrate various important aspects of creating tabular survey data products under a formal privacy guarantee, including utility-risk trade-off. We expect the readers to gain an understanding of how to incorporate survey weights in modeling, how synthetic data works, how to create tabular survey data products once microdata are generated, and the challenges of using Laplace Mechanism.

\section{Introduction}
\label{intro}

%
%
Survey data are collected by government statistical agencies from individuals, households, and business establishments to support research and policy making. 
For example, the Survey of Doctorate Recipients (SDR) provides demographic, education, and career history information from individuals with a U.S. research doctoral degree in a science, engineering, or health (SEH) field. The SDR is sponsored by the National Center for Science and Engineering Statistics and by the National Institutes of Health. Conducted since 1973, the SDR is a unique source of information about the educational and occupational achievements and career movement of U.S.-trained doctoral scientists and engineers in the United States and abroad.

Survey sampling designs typically utilize unequal probabilities for the selection of respondents from the population in order to over-sample important sub-populations or to improve the efficiency of a domain 
(e.g., underrepresented minority)
estimator 
(e.g., of income).
Correlations are also induced in the sampling designs through the sampling of geographic clusters of correlated respondents, which is done for convenience and cost.  As a result of unequal inclusion probabilities and dependence induced by the survey sampling design, the distribution of variables of interest 
(e.g., income) 
are expected to be different in the observed sample than in the underlying population. Therefore, models and statistics estimated on the observed sample without correction will be biased.

At the same time, many government statistical agencies are under legal obligation (such as CIPSEA and Title 13 in the U.S.) to protect the privacy and confidentiality of survey participants.  Agencies utilize statistical disclosure control procedures before releasing any statistics derived from survey responses to the public.  

More recent disclosure limitation methods include the addition of noise to statistics to perturb their values \citep{Dwork:2006:CNS:2180286.2180305}, on the one hand, and the release of synthetic data generated from a model that encodes smoothing to replace the confidential data \citep{Little1993synthetic, Rubin1993synthetic}, on the other hand.  Both classes of methods induce distortion into statistics or data targeted for release to encode privacy protection.  By contrast, survey sampling weights (constructed to be inversely proportional to respondent inclusion probabilities) are used to correct  statistics estimated in the sample to the population.  The goal is to reduce distortion or bias. Such is the case with use of generalized regression estimator (GREG) for producing population statistics from an observed sample \citep{GREG1976}, since survey weights are based on nonresponse adjustments and benchmarking in addition to inclusion probability. Therefore, there is a tension between using survey sampling weights to correct the distortion in the observed sample statistics for the population, on the one hand, from the injection of distortion into data statistics to induce privacy protection, on the other hand. \cite{ShlomoKrenzkeLi2019TDP} is one of the early works that investigates privacy protection approaches for survey weighted frequency tables.

There is a strong connection between smoothness and disclosure risk, where the more local or less smooth is the confidential data distribution, the higher are the identification disclosure risks for survey respondents. The distribution of the sampling weights are typically highly skewed with extremely large values.
In the SDR, individuals are sampled at different rates depending on their doctorate field of study and their demographic information to ensure adequate precision for estimating these domains. As a result, individuals in more common fields and demographic groups are assigned relatively low inclusions probability, which produces higher magnitude sampling weight values.
This skewed distribution for sampling weights can inadvertently accentuate the disclosure risk for a relatively isolated participant
(e.g., someone with an unusual income)
by assigning them a large sampling weight. The result of employing the sampling weights to correct estimates of statistics performed on the observed sample back to the population adds peakedness or roughness to those statistics that, in turn, must be smoothed in a disclosure risk-limiting procedure.

This paper focuses on the adaptation of statistical disclosure control procedures for the production of synthetic data to survey data for estimation of tabular statistics for the population. The utility metrics we consider are tabular cell estimates and associated standard errors of the tabular data constructed from simulated synthetic data with survey weights. In the sequel, we develop two alternatives that correct the survey data distribution to the target population of interest, while simultaneously inducing distortion in the generated synthetic data to reduce the identification disclosure risks for survey respondents. 

\subsection{Differential Privacy}
Our focus metric for measuring the relative privacy guarantees of our additive noise and synthetic data measures is differential privacy (DP) \citep{Dwork:2006:CNS:2180286.2180305}.  Below we provide a formal definition for DP.

\begin{definition}[Differential Privacy]\label{def:DP}
Let $D \in \mathbb{R}^{n \times k}$ be a database in input space $\mathcal{D}$. Let $\mathcal{M}$ be a randomized mechanism such
that $\mathcal{M}(): \mathbb{R}^{n \times k} \rightarrow O$. Then $\mathcal{M}$ is $\epsilon$-differentially private if
\[
\frac{Pr[\mathcal{M}(D) \in O]}{Pr[\mathcal{M}(D^{'}) \in O]} \le \exp(\epsilon),
\]
for all possible outputs $O = Range(\mathcal{M})$ under all possible pairs of datasets $D, D^{'} \in \mathcal{D}$ of the same size which differ by only a single row (Hamming-1 distance).
\end{definition}

DP assigns a disclosure risk for a statistic to be released to the public, $f(D)$ (e.g., total employment for a state-industry) of any $D \in \mathcal{D}$ based on the global sensitivity, $\Delta_G = \mathop{\sup}_{D,D^{'}\in\mathcal{D}: ~\delta(D,D^{'})=1}\abs{f(D) - f(D^{'})}$, over the space of databases, $\mathcal{D}$, where $\delta(D,D^{'}) = 1$ denotes the Hamming$-1$ distance such that $D$ differs from $D^{'}$ by a single record. If the value of the statistic, $f$, expresses a high magnitude change after the change of a data record from $D^{'}$, then the mechanism will be required to induce a relatively higher level of distortion to $f$.  The more sensitive is a statistic to the change of a record, the higher its disclosure risk. 

Under additive noise processes, such as the Laplace Mechanism \citep{Dwork:2006:CNS:2180286.2180305}, $\mathcal{M}(D) = f(D) + \eta$ with $E(\eta) = 0$.
The scale of additive noise to produce a DP guarantee of $\epsilon$ is proportional to $\Delta_G/\epsilon$, where the larger the sensitivity, $\Delta_G$, the higher the required scale for the addition of Laplace-distributed noise to $f$. 
The DP guarantee, $\epsilon$, is a property of the randomized mechanism $\mathcal{M}$, not the actual released data $\mathcal{M}(D)$.  The privacy guarantee, $\epsilon$, may be viewed as a budget \citep{Dwork:2006:CNS:2180286.2180305} that may be expended on selective releases of privacy-protected statistics.  An example is a mechanism that outputs a randomized statistic, $f(q_{j}(D))$, for query, $q_{j}$, by adding Laplace noise proportional to $\Delta_{G,j}/\epsilon_{j}$, equipped with a privacy guarantee, $\epsilon_{j}$. $\Delta_{G,j}$ denotes the global sensitivity over the space of databases, $\mathcal{D}$, of the randomized query statistic, $f(q_{j})$. If there are $J$ such queries, then the owner of the confidential data will set each $\epsilon_{j}$ such that an overall target guarantee, $\epsilon = \sum_{j=1}^{J}\epsilon_{j}$ is achieved.  Under this setup, the $\epsilon$ guarantee is viewed as a budget that is allocated to account for the information disclosed in each query.

\subsection{Synthetic Data}

An alternative to the addition of additive noise to statistics is the generation of synthetic data to replace the confidential data for release to the public.  Synthetic data is produced by estimating a model on the confidential data, followed by generating replicate data from the model posterior predictive distribution \citep{Little1993synthetic, Rubin1993synthetic}. A major advantage of synthetic data methods, in contrast with additive noise mechanisms, is that they do not need to account for interactive query data releases that spend a privacy budget for each release mechanism computed on the confidential data. The synthetic data release comes with a certain level of privacy protection. Moreover, these data may be used for any purpose, including producing unlimited tables with cells at any level of granularity, although their utility is often excellent on predetermined analysis-specific measures but not necessarily other measures \citep{Snoke2018JRSSA}. Nevertheless, from the privacy protection perspective, there is no subsequent privacy ``accounting” required after the initial creation of the synthetic data.

\citet{Dimitrakakis:2017:DPB:3122009.3122020} employ the Exponential Mechanism of \citet{McSherryTalwar2007} for generating synthetic data by selecting the model log-likelihood as the utility function, which produces the posterior distribution, $\xi(\bm{\theta} \mid \bm{X})$, as the random mechanism, $\mathcal{M}(\bm{X},\bm{\theta})$.  They demonstrate a connection between the model-indexed sensitivity, $ \mathop{\sup}_{\bm{x},\bm{y}\in \mathcal{X}^{n}:\delta(\bm{x}, \bm{y}) = 1}  \mathop{\sup}_{\theta \in \Theta} \lvert f_{\theta}(\bm{x}) - f_{\theta}(\bm{y}) \rvert \le \Delta$ and $\epsilon \leq 2\Delta$, where $f_{\theta}(\bm{x})$ is the model log-likelihood and $\Delta$ denotes a Lipschitz bound.  The guarantee applies to all databases $\bm{x}$ in the space of databases of size $n$, $\mathcal{X}^{n}$.  The posterior mechanism suffers from a non-finite $\Delta = \infty$ for most Bayesian probability models used, in practice. 

\subsection{Survey Data}

Suppose a sample $S$ of $n$ individuals is taken from a population $U$ of size $N$.
The sample is taken under a survey design distribution that assigns indicators $\omega_i \in \{0,1\}$ to each individual in $U$ with probability of selection $P(\omega_i = 1 \mid \mathcal{A}) = \pi_i$, where $\mathcal{A}$ denotes the accumulated information in the population. Often the selection probabilities $\pi_i$ are related to the response of interest $y_i$. The balance of information of the observed sample $y_i,~i \in S$ is different from the balance of information in the population $y_\ell,~ \ell \in U$, commonly known as an \emph{informative} sampling design \citep{PfeffermannSverchkov2009chapter}.
To account for this imbalance, survey weights $w_i = 1/\pi_i$ are used to create estimators on the observed sample that reduce bias; for example, a consistent estimator of the population mean is
$\hat{\mu} = \sum_{S} w_i y_i / \sum_{S} w_i$. When the estimation focus moves beyond simple statistics, consistent estimation for more general models can be based on the exponentiated pseudo likelihood: $L^{\bm w}(\theta) = \prod_{S}p(y_i|\theta)^{w_i}$. Use of this pseudo likelihood in Bayesian probability models, which we utilize in the sequel, provides for consistent estimation of $\theta$ for broad classes of both population models \citep{Savitsky16} and complex survey sampling designs \citep{williams2020}.  \citet{10.2307/1403139} and \citet{doi:10.1177/096228029600500303} discuss use of the pseudo likelihood in frequentist estimation.

The use of survey weights increases the influence of individual observations, thus increasing the sensitivity of the output mechanism. For example, a weighted count would have a sensitivity of $\max_{i} w_{i}$ instead of 1 in an unweighted case. Thus direct usage of additive noise and perturbation can lead to a large amount of noise at the expense of utility.

For estimation, survey weights mitigate the estimation bias, but the uncertainty distribution (covariance structure) also needs to be estimated and adjusted \citep{williams2018bayesian}.
The typical assumption for variance estimation (of the same models) is to assume an arbitrary amount of within cluster dependence both in the sampling design and the population generating model \citep{heeringa2010applied, Rao92}.

The de facto approach for variance estimation is based on the approximate sampling independence of the primary sampling units \citep{heeringa2010applied}. Variance estimation can be in the form of Taylor linearization or replication based methods with a variety of implementations available for each \citep{binder1996, Rao92}. \citet{williams2018bayesian} propose a hybrid approach made possible by recent advances in algorithmic differentiation \citep{Margossian18}. Each of these methods re-use the data and thus require additional privacy budget allocation. Trade-offs between efficient estimation of variance (full number of clusters or replicates for the full precision of variance estimates) and conserving an $\epsilon$ budget (aggregating clusters and reducing replicates for reduced precision of variance estimates) are an open challenge.

\subsection{Pseudo Posterior Mechanism}
\label{intro:main}

In this work, we aim at producing tabular data products with privacy guarantee $\epsilon$, conditioned on the local survey sample database. Specifically, we propose to synthesize microdata under the pseudo posterior mechanism \citep{SavitskyWilliamsHu2020ppm}, coupled with survey weights. In other words, we extend \citet{SavitskyWilliamsHu2020ppm} to generating synthetic survey microdata. We may view the focus on tabular data, formed from the synthetic survey microdata, as one type of data utility for evaluating our adapted pseudo posterior mechanism.  We note that, once generated, synthetic survey microdata may be used for many purposes, each with distinct measures of utility, e.g., weighted regressions with survey weights. Our focus on tabular statistics owes to their being the main data product released by government statistical agencies.  Motivated by the SDR application presented in Section \ref{app}, which stratifies by field of study and oversamples based on gender and underrepresented minority status, we utilize an informative single stage stratified sample design with unequal probabilities of selection (with or without replacement).

We consider a local survey database $\bm y_n = (y_1, \cdots, y_n)$, design information in variables $\bm X_n = (\bm x_1, \cdots, \bm x_n)$, and associated survey weights $\bm w_n = (w_1, \cdots, w_n)$. We consider univariate continuous outcome variable $y_i$, such as salary, and categorical design information vector $\bm x_i$, such as field of study and gender, where fields of study are strata. These design variables are considered public information in our setup. Our goals are to create private tables of counts (cell, marginal, and total) and average salary (cell, marginal, and total) by field and gender from the synthetic microdata containing synthesized $\bm y_n$ and $\bm w_n$, with $\epsilon$ privacy guarantee.

\subsubsection{Review of the pseudo posterior mechanism}
\label{intro:main:review}
The basic setup of our proposed survey microdata synthesizers uses privacy protection weights $\bm \alpha$ under the pseudo posterior mechanism. This mechanism estimates an $\bm{\alpha}-$weighted pseudo posterior distribution (without survey weights $\bm w_n$),
\begin{equation}
\label{eq:pp}
\xi^{\bm{\alpha}(\bm{y}_n)}(\bm \theta \mid \bm{y}_n, \bm X_n) \propto \mathop{\prod}_{i=1}^{n}p(y_{i}\mid \bm x_i, \bm \theta)^{\alpha_{i}} \times \xi(\bm \theta), 
\end{equation}
where $\xi(\bm \theta)$ is the prior distribution and $p(\cdot)$ denotes the likelihood function with corresponding utility $\sum_{i=1}^{n} \alpha_{i}u(y_i, \bm x_i, \theta) = \log\left( \mathop{\prod}_{i=1}^{n}p(y_{i}\mid \bm x_i, \theta)^{\alpha_{i}}\right)$ that generalizes the Exponential Mechanism. Define $\Delta_{\bm{\alpha}}$ as the $\bm{\alpha}-$weighted Lipschitz bound or sensitivity over the space of databases and the space of parameters. 

The weights, $(\alpha(y_{i}))$, are formulated \emph{locally} from the observed database, $\bm{y}_{n}$.  We denote the Lipschitz bound computed locally from the observed database as $\Delta_{\bm{\alpha},\bm{y}_{n}}$.  Differential privacy is a guarantee over the space of all databases, $\bm{y}_{n} \in \mathcal{Y}^{n}$ of size $n$.  \citet{SavitskyWilliamsHu2020ppm} demonstrate that the pseudo posterior mechanism of Equation~(\ref{eq:pp}) satisfies a formal guarantee that they label ``asymptotic differential privacy" with level $\epsilon = 2\Delta_{\bm{\alpha}}$, or aDP$-\epsilon$.  Asymptotic differential privacy guarantees that the Lipschitz bounds computed locally on a collection of observed databases, $\Delta_{\bm{\alpha},\bm{y}_{n}}$, contract onto a global Lipschitz bound, $\Delta_{\bm{\alpha}}$, as the number of database observations, $n$, increases.   The contraction of the local Lipschitz bounds onto the global bound occurs at $\mathcal{O}(n^{-1/2})$.  We may imagine the generation of multiple collections of databases, $\{\bm{y}_{n,r}\}_{r=1}^{R}$, that produce the associated collection of Lipschitz bounds, $(\Delta_{\bm{\alpha},\bm{y}_{n,r}})_{r=1}^{R}$.  The aDP result guarantees that for $n$ sufficiently large, the local Lipschitz bounds for that collection of databases contract onto the global Lipschitz bound. In practice, \citet{SavitskyWilliamsHu2020ppm} show that in a Monte Carlo simulation study that for sample sizes of a few hundred that the $\Delta_{\bm{\alpha},\bm{y}_{n}} = \Delta_{\bm{\alpha}}$ up to any desired precision.

The details of steps of calculating $\alpha_i \propto 1 / \Delta_{\bm \alpha, y_i} \in [0,1]$ based on the local database $(\bm y_n, \bm X_n)$ are laid out in Algorithm 1 in \cite{SavitskyHuWilliams2020reweight}.  A quick overview of the algorithm, including the generation of synthetic data, involves the following steps:
\begin{enumerate}
    \item Estimate the unweighted posterior distribution, $\xi(\bm \theta \mid \bm{y}_n, \bm X_n)$, and obtain parameter draws from the model posterior distribution for $\bm \theta$.
    \item Compute record-indexed weight, $\alpha_{i} \propto 1 / \displaystyle\mathop{\max}_{s \in (1,\ldots,S)}\abs{\log(p(y_{i} \mid \bm x_{i},\theta_{s})} = 1/\Delta_{y_{i}}$, where $s \in (1,\ldots,S)$ represents a draw of the posterior distribution from the MCMC estimation algorithm used for the model and $\Delta_{y_{i}}$ is the Lipschitz bond local to database, $\bm{y}_{n}$, when all weights $\bm{\alpha}$ are set to $1$.  The maximum value of the absolute value of the log likelihood for record $i$ over the parameter space represents the disclosure risk for that record. A high risk data record tends to be isolated from other data records such as in the case of a very high income individual where the sensitive variable of focus are incomes for a collection of individuals.  In the case that the data value for a record is isolated, its absolute log likelihood value will be very large since there is little probability mass placed in that tail region of the distribution and the resulting $\alpha_{i}$ for that record will be relatively small. Each record-indexed weight, $\alpha_{i}$, is scaled to lie in $[0,1]$, where the likelihood contribution for a relatively higher risk record is downweighted such that the data for that record, $y_{i}$, is partly excluded from estimation of the pseudo posterior model of Equation~(\ref{eq:pp}).  A record with a non-finite risk is assigned weight, $\alpha_{i} = 0$, and is completely excluded from the pseudo posterior estimation model.
    \item Use the computed $\bm\alpha(\bm y_n) = (\alpha_{1},\ldots,\alpha_{n})$ to estimate the pseudo posterior distribution for $\bm\theta$ in Equation~(\ref{eq:pp}) and compute the record-level Lipschitz, $\displaystyle\Delta_{\bm{\alpha},y_{i}} = \mathop{\max}_{s\in (1,\ldots,S)}\abs{\alpha_{i}\times \log(p(y_{i} \mid \bm x_{i},\theta_{s})} \leq \Delta_{y_{i}}$, where the latter inequality derives from $\alpha_{i} \leq 1$ for all $i \in (1,\ldots,n)$.
    \item Compute the overall Lipschitz bound of the pseudo posterior mechanism local to database $\bm{y}_{n}$, $\displaystyle\Delta_{\bm{\alpha}, \bm y_n} = \mathop{\max}_{i \in (1,\ldots,n)} \Delta_{\bm{\alpha},y_{i}}$. As contrasted with the posterior mechanism of \citet{Dimitrakakis:2017:DPB:3122009.3122020}, the pseudo posterior mechanism guarantees $\Delta_{\bm{\alpha},y_{n}} < \infty$ by setting the $\alpha_{i} = 0$ for any record with a non-finite absolute log-likelihood. 
    \item With a set of parameter draws for $\bm \theta^{(\ell)}$, simulate a synthetic dataset $\bm y_n^{*, (\ell)}$ according to the sampling model $(y_i \mid \bm x_i, \bm \theta)$. If $m$ synthetic datasets are simulated ($\ell = 1, \cdots, m$), the synthetic data release determines the privacy guarantee, $\epsilon_{\bm y_n} = 2\Delta_{\bm{\alpha}, \bm y_n} \times m$.  
\end{enumerate}

There is no leakage in computation of the data-dependent weights, $\bm{\alpha}(\bm y_n)$, to formulate the pseudo posterior mechanism, because these privacy protection weights are not released, but serve to downweight or remove the data contributions of highly risky records.  The overall sensitivity and privacy guarantee are released, as is the synthetic microdata generated from the pseudo posterior mechanism (via the model posterior predictive distribution).  The lower the weights are, $(\alpha_{i})_{i=1}^{n}$, the greater the degree of distortion that will be induced in the distribution of the synthetic data as compared to the confidential data.  It bears mention that the each weight, $\alpha_i$, can be scaled and shifted by $(c_1, c_2)$ to $\alpha^{\ast} \in [0, 1]$ (as in $\alpha^{\ast}_{i} = c_1 \times \alpha_{i} + c_2$), to increase or decrease the weights as an \emph{indirect} means to achieve a target $\epsilon_{\bm y_n} = 2\Delta_{\bm{\alpha}^{\ast}, \bm y_n}$ for generation of synthetic databases. 

The pseudo posterior mechanism has the virtue that it may be applied to \emph{any} data synthesizing model and is readily estimable while being equipped with an asymptotic DP guarantee.

\subsubsection{Including survey weights}

The pseudo posterior mechanism allows us to take a synthesizer model with good utility and modify it to become asymptotically differentially private. For survey data with sample weights, we could jointly synthesize the response variable $\bm y_n$ together with the survey weights and then use the synthesized weights to extrapolate the synthesized response variable back to the population. Alternatively, we could directly synthesize from the population distribution for the response variable by using the survey weights released with the observed sample to correct back to the population during estimation. The former method synthesizes both the outcome variable $\bm y_n$ and weights $\bm w_n$ under the distribution of the observed sample and then uses the synthesized weights to correct the data back to the population distribution. The latter method uses the survey weights to synthesize the outcome variable under the population distribution and then discards the weights. In either approach, the design variables $\bm X_n$ are considered public. We note that if any of these design variables were deemed private, we would co-model and synthesize them without loss of generality.

%


In this work, we propose two microdata synthesizers that incorporate survey weights under the pseudo posterior mechanism framework: (i) A Fully Bayes model for observed sample models $(\bm y_n, \bm w_n)$ under a bivariate normal model on the transformed data, using privacy protection weights $\bm \alpha$ in the joint likelihood, and the public $\bm X_n$ are used as predictors and unsynthesized; (ii) A Fully Bayes model for the population that forms the exact likelihood for $(\bm y_n, \bm w_n)$ in the observed sample to model $\bm y_n$ and corrects for population bias \citep{pkr:1998, LeonSavitsky2019EJS}, where privacy protection weights $\bm \alpha$ are used and the public $\bm X_n$ are used as predictors and unsynthesized.  
Both methods are scaled to have equivalent asymptotic differential privacy guarantee $(\epsilon_{\bm y_n} = 2 \Delta_{\bm \alpha, (\bm y_n, \bm X_n, \bm w_n)} \times m)$, where $m \geq 1$ is the number of simulated synthetic datasets. Once synthetic microdata are available from each approach, we create survey tables of counts and average outcome by design variables and compare their utility performances, mainly the point estimates and standard error estimates.



We utilize a comparison method that adds noise proportional to the sensitivity local to the database from the Laplace Mechanism. A detailed review of the Laplace Mechanism is available in the Supplementary Materials. We use the same level of privacy guarantee $(\epsilon_{\bm y_n} = 2 \Delta_{\bm \alpha, (\bm y_n, \bm X_n, \bm w_n)} \times m)$ as in our two microdata synthesis approaches. With the same level of privacy guarantee, we compare their utility performances in the created survey tables of counts and average salary by design variables. The privacy guarantee is ``local" for the Laplace Mechanism in that is applies to the single observed dataset rather than the collection of all possible data sets of the same type (i.e., global). The pseudo posterior mechanism results, by contrast, are asymptotically (globally) differentially private.  The comparisons between the two mechanisms on an observed or local database are useful to assess differences in utility performance for the same privacy guarantee, even though the pseudo posterior mechanism is equipped with an asymptotic DP guarantee.

The remainder of the paper is organized as follows. In Section \ref{methods}, we lay out the details of our two proposed microdata synthesis approaches, with a discussion and comparison between the two, as well as how to create survey tables from synthetic microdata in each approach. We also describe details of how to construct the sensitivity of the Laplace Mechanism with survey weights. We present a real data application to a sample of the SDR in Section \ref{app}, where we focus on utility comparison among the three methods. The application results motivate us to create a series of simulation studies under informative sampling design in Section \ref{sim}, where we compare the utility performances, investigate the tuning of the utility-risk trade-off of our two microdata synthesis approaches, and their performances under repeated sampling. The paper ends with concluding remarks in Section \ref{conclusion}. 

\section{Methods for Differentially Private Tabular Survey Data}
\label{methods}

In this section, we describe the fully Bayes model for observed sample in Section \ref{methods:FBS} and the fully Bayes model for the population in Section \ref{methods:FBP}. We discuss and compare these two approaches in Section \ref{methods:comparison}, where we also describe in detail how to create survey tables from synthetic microdata under each approach. Finally, we present how to add noise from the Laplace Mechanism with survey weights in Section \ref{methods:Dwork}.

\subsection{Fully Bayes Model for Observed Sample (FBS)}
\label{methods:FBS}


Our first approach is fully Bayesian because it jointly models the outcome $\bm y_n$ and sampling weight $\bm w_n$ of the observed sample of size $n$. We label it FBS (Fully Bayes Sample). Since we model the observed sample, not the population, we do not assume the model estimated on the sample is the population generating model.  In fact, the distribution of the outcome and weight variables are generally expected to be different in the observed sample than for the underlying population.  We retain the smoothed / model-estimated version of the outcome and weights, and we utilize the latter to correct the distribution of the outcome in the sample back to the population.  We use a bivariate normal synthesizer for the joint distribution of $(\tilde{y}_i, \tilde{w}_i)$ of unit $i$, where $\tilde{y}_i$ and $\tilde{w}_i$ are $y_i$ and $w_i$ after appropriate transformation (e.g., log transformation) with predictors $\bm x_i$, as in Equation~(\ref{eq:FBSmodel}):

\begin{equation}
\begin{bmatrix}
\tilde{y}_i \\
\tilde{w}_i
\end{bmatrix}
\sim \textrm{MVN}_2(\bm x_i \bm \beta, \Sigma) = \textrm{MVN}_2\left(\bm x_i 
\begin{bmatrix}
\bm \beta_y \\
\bm \beta_w
\end{bmatrix}, \Sigma\right).
\label{eq:FBSmodel}
\end{equation}
For synthesizing data ($y_i, w_i$), we then back-transform the generated MVN values of ($\tilde{y}_i, \tilde{w}_i$).
We specify an independent and identically-distributed multivariate Gaussian prior for the coefficient locations, $(\boldsymbol{\beta}_y)$ and $(\boldsymbol{\beta}_w)$, and a uniform prior for covariance matrix, $\Sigma$, over the space of covariance matrices of size $2 \times 2$  \citep{Rstan}. The details of the prior specification are included in the Supplementary Materials.


We fit this unweighted synthesizer for $(\bm y_n, \bm X_n, \bm w_n$) and calculating the unit-level privacy protection weights $\bm \alpha = (\alpha_1, \cdots, \alpha_n)$ using the procedure mentioned in Section \ref{intro:main:review} where we use Stan \citep{Rstan} to provide posterior estimates for parameters, $\bm\theta = (\bm\beta,\Sigma)$. For each unit $i$, we exponentiate its likelihood by $\alpha_i$, so that we arrive at the $\bm \alpha-$weighted pseudo posterior distribution of $\bm\theta = (\bm \beta, \Sigma)$, as in Equation~(\ref{eq:FBSpseudo}):

\begin{equation}
\xi^{\bm{\alpha}(\bm y_n, \bm X_n, \bm w_n)}(\bm \beta, \Sigma \mid \bm y_n, \bm X_n, \bm w_n) \propto \mathop{\prod}_{i=1}^{n}p(y_i, w_i \mid  \bm x_i, \bm \beta, \Sigma)^{\alpha_{i}} \times \xi(\bm \beta, \Sigma).
\label{eq:FBSpseudo}
\end{equation}

Once we estimate this $\bm \alpha-$weighted pseudo posterior distribution of $(\bm \beta, \Sigma)$, we simulate $m$ posterior samples, which achieve the local $(\epsilon_{\bm y_n} = 2 \Delta_{\bm \alpha, (\bm y_n, \bm X_n, \bm w_n)} \times m)$ privacy guarantee. Given the simulated $m$ posterior samples of $(\bm \beta, \Sigma)$, we can generate $m$ synthetic survey datasets following the bivariate normal model in Equation~(\ref{eq:FBSmodel}), denoted as $(\bm Y^*, \bm X, \bm W^*) = \{(\bm y_n^{*, (1)}, \bm X_n^{(1)}, \bm w_n^{*, (1)}), \cdots, (\bm y_n^{*, (m)}, \bm X_n^{(m)}, \bm w_n^{*, (m)})\}$, where superscript $^*$ refers to synthetic.

Each synthetic survey dataset $(\bm y_n^{*, (\ell)}, \bm X_n^{(\ell)}, \bm w_n^{*, (\ell)})$, $\ell = 1, \cdots, m$, is used to form survey tables, which are to be released. This table creation process does not cost additional privacy budget since it is post-processing (the protected data $\bm y_n$ is not used; \citet{Dwork:2006:CNS:2180286.2180305, NissimRaskhodnikovaSmith2007ACM}). Moreover, since the predictors $\bm x_i$ are not synthesized, this approach produces partially synthetic data, and the survey tables are created by combining rules of partial synthesis \citep{ReiterRaghu2007, Drechsler2011book}. Details of the combining rules are included in Appendix \ref{sec:appendix:combiningrules}.


In addition, we use smoothed $\bm w_n^{*, (\ell)}$ from the conditional normal distribution derived from Equation~(\ref{eq:FBSmodel}), $E(w_i \mid y_i) = \bm x_i \bm \beta_w +  \rho (y_i - \bm x_i \bm \beta_y) \sigma_w / \sigma_y$ (where $\rho$ denotes the correlation between $\bm y_n$ and $\bm w_n$; $\sigma_y$ and $\sigma_w$ are the standard deviation of $\bm y_n$ and $\bm w_n$, respectively) to create survey tables from synthetic survey sample $(\bm y_n^{*, (\ell)}, \bm X_n^{(\ell)}, \bm w_n^{*, (\ell)})$. The smoothed weights $\bm w_n^{*, (\ell)}$ will provide survey tables with less noise, an appealing feature of modeling $\bm w_n$ with outcome variable $\bm y_n$.


We can also directly release synthetic survey samples $(\bm Y^*, \bm X, \bm W^*)$ to the public. Data users will need to know how to incorporate survey weights $\bm w_n^{*, (\ell)}$ for unbiased inference with respect to the population. The availability of synthetic survey samples allows data users to perform analyses of their interests which are only feasible with microdata, such as a weighted regression of income $\bm y_n$ on predictors $\bm X_n$, whose accuracy needs to be evaluated. Nevertheless, this increases the utility of the synthetic data compared to direct table protection without additional privacy loss.

\subsection{Fully Bayes Model for Population (FBP)}
\label{methods:FBP}

Our second approach is also fully Bayesian and jointly models the outcome $\bm y_n$ and the sampling weights $\bm w_n$. However, the specific joint specification modeled (on the observed sample) is for the generative model of the population and the sample design, rather than directly modelling the sample itself. We label it FBP (Fully Bayes Population). To form the \emph{exact} likelihood for $(y_i, w_i)$ in the observed sample which corrects for population bias, we follow the fully Bayesian approach proposed by \citet{LeonSavitsky2019EJS} and form the exact likelihood through inclusion probability $w_i = 1 / \pi_i$; that is, we model $(y_i, \pi_i)$ in the observed sample. We first assume a linear model for the population as
\begin{equation}
y_i \mid \bm x_i, \bm \beta, \sigma_y^2 \sim \textrm{Normal}(\bm x_i^t \bm \beta, \sigma_y^2).
\label{eq:FBPmodel1}
\end{equation}

Given $y_i$, the conditional population model for inclusion probabilities is
\begin{equation}
\pi_i \mid y_i, \bm x_i, \kappa_y, \bm \kappa_x, \sigma_{\pi}^2 \sim \textrm{Lognormal}(\kappa_y y_i + \bm x_i^t \bm \kappa_x, \sigma_{\pi}^2),
\label{eq:FBPmodel2}
\end{equation}
where $\kappa_y$ and $\bm \kappa_x$ are regression coefficients for $y_i$ and $\bm x_i^t$, respectively. We specify independent multivariate normal priors for $\bm \beta$ and $(\kappa_y, \bm \kappa_x)$ and half Cauchy priors for $\sigma_y$ and $\sigma_x$. The details of the prior specification are available in the Supplementary Materials. In practice, we may first transform the outcome $(\log(y_i))$.

\citet{LeonSavitsky2019EJS} has shown that the posterior distribution observed on the sample is 
\begin{eqnarray}
\xi_s(y_i, \pi_i \mid \bm x_i, \bm \beta, \sigma_y, \kappa_y, \bm \kappa_x, \sigma_{\pi}) &\propto& \frac{\textrm{Normal}(\log \pi_i \mid \kappa_y y_i + \bm x_i^t \bm \kappa_x, \sigma_{\pi}^2)}{\exp \{\bm x_i^t \bm \kappa_x + \sigma_{\pi}^{2} / 2 + \kappa_y \bm x_i^t \bm \beta + \kappa_y^2 \sigma_y^{2} / 2\}} \nonumber \\
&&\times \textrm{Normal}(y_i \mid \bm x_i \bm \beta, \sigma_y^2).
\end{eqnarray}


After fitting this unweighted synthesizer for $(\bm y_n, \bm X_n, \bm \pi_n)$, we calculate the unit-level privacy protection weights $\bm \alpha = (\alpha_1, \cdots, \alpha_n)$. For each unit $i$, we exponentiate its likelihood by $\alpha_i$, so that we arrive at the $\bm \alpha-$weighted pseudo posterior distribution of $( \bm \beta, \sigma_y, \kappa_y, \bm \kappa_x, \sigma_{\pi})$, as in Equation~(\ref{eq:FBPpseudo}):
\begin{eqnarray}
\xi^{\bm{\alpha}(\bm y_n, \bm X_n, \bm \pi_n)}( \bm \beta, \sigma_y, \kappa_y, \bm \kappa_x, \sigma_{\pi} \mid \bm y_n, \bm \pi_n, \bm X_n) &\propto& \mathop{\prod}_{i=1}^{n}p(y_i, \pi_i \mid  \bm x_i,  \bm \beta, \sigma_y, \kappa_y, \bm \kappa_x, \sigma_{\pi})^{\alpha_{i}} \nonumber \\
&& \times \xi(\bm \beta, \sigma_y, \kappa_y, \bm \kappa_x, \sigma_{\pi}).
\label{eq:FBPpseudo}
\end{eqnarray}

Once we estimate this $\bm \alpha-$weighted pseudo posterior distribution of $(\bm \beta, \sigma_y, \kappa_y, \bm \kappa_x, \sigma_{\pi})$, we simulate $m$ posterior samples, which achieve the local $(\epsilon_{\bm y_n} = 2 \Delta_{\bm \alpha, (\bm y_n, \bm X_n, \bm w_n)} \times m)$ privacy guarantee\footnote{Note that for coherence with the FBS approach, we use $\bm w_n$ instead of $\bm \pi_n$ in the expression of $\epsilon_{\bm y_n}$, and this can be done since $w_i \propto 1 / \pi_i$.}. Given the simulated $m$ posterior samples of $(\bm \beta, \sigma_y, \kappa_y, \bm \kappa_x, \sigma_{\pi})$, we can generate $m$ synthetic survey datasets following the population model in Equation~(\ref{eq:FBPmodel1}) and Equation~(\ref{eq:FBPmodel2}), denoted as $(\bm Y^*, \bm X, \bm W^*) = \{(\bm y_n^{*, (1)}, \bm X_n^{(1)}, \bm w_n^{*, (1)}), \cdots, (\bm y_n^{*, (m)}, \bm X_n^{(m)}, \bm w_n^{*, (m)})\}$, where each $w_i^{*, (l)} \propto 1 / \pi_i^{*, (l)}$, and $\ell = 1, \cdots, m$ (the $\propto$ is used to account for normalization).

Similar to FBS, each synthetic survey dataset $(\bm y_n^{*, (\ell)}, \bm w_n^{*, (\ell)})$, $\ell = 1, \cdots, m$, is used to form survey tables. Partial synthesis combining rules are used to create the survey tables due to the fact that the predictors $\bm x_i$ are not synthesized. As with FBS, the table creation process does not cost additional privacy budget. Moreover, smoothed weights $\bm w_n^{*, (\ell)}$ are generated from the smoothed $\bm \pi_n^{*, (\ell)}$ by using only the $\kappa_y y_i$ component of the mean in Equation~(\ref{eq:FBPmodel2}). These weights are only needed when combining data across strata. Within strata, the individuals in the synthetic population are equally weighted. See Section \ref{methods:comparison} for more details.

Alternatively, we can directly release synthetic survey samples $(\bm Y^*, \bm X, \bm W^*)$ to the public. Data users do not need to know how to incorporate survey weights $\bm w_n^{*, (\ell)}$ for unbiased inference with respect to the population, since $\bm Y^*$ are corrected for survey sampling bias. However, data users need to aggregate survey weights $\bm w_n^{*, (\ell)}$ to account for differences in population sizes across strata, for example when creating tables of counts. As with FBS, the release of synthetic survey samples $(\bm Y^*, \bm X, \bm W^*)$ increases the utility of the synthetic data compared to direct table protection without additional privacy loss.

\subsection{Comparison and Discussion of Two Synthesis Approaches}
\label{methods:comparison}

In this section, we discuss and compare the two microdata synthesis approaches, whose key features are summarized and compared in Table \ref{tab:2approaches}. 

\begin{table}[H]
\centering
\begin{tabular}{ l | c c }
\hline
 & FBS & FBP  \\ \hline
Modeling $\bm w_n$ & yes & yes \\ 
Bias correction stage & analysis & synthesis/analysis  \\
Synthesis type & partial synthesis & partial synthesis  \\
Synthesized variables & $(\bm y_n, \bm w_n)$ & $(\bm y_n, \bm w_n)$  \\
Unsynthesized variables & $\bm X_n$ &  $\bm X_n$ \\ 
$m$ synthetic samples & $(\bm Y^*, \bm X, \bm W^*)$ & $(\bm Y^*, \bm X, \bm W^*)$ \\
Post analysis & weighted/stratified & stratified  \\
\hline
\end{tabular}
\caption{Features of the two approaches: Fully Bayes for observed Sample (FBS) and Fully Bayes for Population (FBP).}
\label{tab:2approaches}
\end{table}

Both approaches are fully Bayesian approaches which jointly model outcome variables and the sampling weights. FBS models the observed sample without correction for population bias, therefore we will use the synthesized sampling weights $\bm w^*_n$ to form survey tables (correcting the bias at the analysis stage). 
FBP corrects for bias at the modeling stage. However, since FBP does not co-model $\bm X_n$ ($\bm X_n$ are used as predictors), the synthetic weights will be used to generate marginal estimates across different values of $\bm X_n$. In other words, the partially synthetic FBP does not correct for the difference in the distributions of $\bm X_n$ between the sample and the population. Both approaches implement partial synthesis because design variables $\bm X_n$ are used as predictors but not synthesized. We also note that while FBP can incorporate population bias correction directly into the model, it is less flexible compared to FBS in more complicated settings, for example, when there are more than one outcome variables of mixed types.

An important aspect of discussing and comparing the two approaches is how to create survey tables from each approach once synthetic microdata are obtained. These tables include both point estimates and standard error estimates.

FBS requires correction for population bias of the outcome variable with survey weights. Therefore, with the $\ell$th ($\ell = 1, \cdots, m$) synthetic sample from FBS, $(\bm y_n^{*, (\ell)}, \bm X_n^{(\ell)}, \bm w_n^{*, (\ell)})$, we use the smoothed weights $\bm w_n^{*, (\ell)}$ to create the counts and average salary values by domain variables (field $f$ and gender $g$):
$\hat{N}_{fg} = \sum_{i \in S} \mathbbm{1}^{i}_{fg} w^{*}_i $ and
$\hat{\mu}_{fg} = (\sum_{i \in S} \mathbbm{1}^{i}_{fg} w^{*}_i y^{*}_i)/\hat{N}_{fg}$, where $\mathbbm{1}^{i}_{fg} \in \{0,1\}$ is an indicator variable for an individual $i$ belonging to cell $\{f,g\}$.
Variance estimates for a cell total count and salary for \emph{each} database, $\ell$, are produced via Taylor linearization \citep{binder1996}, a standard method for survey samples. For linear estimators such as totals and means, the approach is straight forward:
\begin{itemize}
    \item For each cluster $c$ of the $n_c$ sampled clusters, define the aggregate residual $r_{c} = \sum_{i \in c} \mathbbm{1}^{i}_{fg} (w^{*}_i y^{*}_i/\hat{N}_{fg} - \hat{\mu}_{fg})$. Note that $\sum_{c \in S} r_c = 0$ exactly for linear estimators.
    \item Estimate variance $\widehat{Var(\hat{\mu}_{fg})} = \widehat{Var(r_c)} = \frac{1}{n_c - 1} \sum_{c \in S}(r_{c} - \bar{r_{c}})^2$ with $\bar{r_{c}} = \frac{1}{n_c} \sum_{c \in S}r_{c} = 0$.
\end{itemize}

For a one-stage stratified design, such as the SDR, each respondent is an individual unit ($i = c$) and variance calculations are performed independently across strata and then aggregated. 
The $m$ synthetic databases are used to compute a between-databases variance and combining rules of partial synthesis are used to compute final point and standard error estimates. 

FBP incorporates correction for population bias of the outcome variable in the model for given values of $\bm X_n$. In our example, $\bm X_n$ are categorical data. Therefore, we create average salary values  $\hat{\mu}_{fg}$ by design variables  \emph{without} any weights $(w^{*} = 1)$.  Because we do not synthesize $\bm X_n$, however, we use the smoothed weights, $\bm w_n^{*, (\ell)}$, to construct marginal salary values (e.g., over gender $\hat{\mu}_{g}$ and field $\hat{\mu}_{f}$). When creating counts (or size) estimates $\hat{N}_{fg}$ for the population based on the $\bm X_n$ categories, we create counts \emph{with} smoothed weights $\bm w_n^{*, (\ell)}$. As with FBS, we use combining rules of partial synthesis to create final point and standard error estimates. 
We estimate variances for the tabular data using the same Taylor linearization approach outlines above for FBS.


With co-modeling of survey weights together with the outcome salary and using smoothed weights in table construction, FBS and FBP will result in more accurate point estimates and smaller standard error estimates of counts and average salary values than any mechanism which uses the raw (unsmoothed) sampling weights, such as additive noise mechanisms. In the sequel, we ensure robust Markov chain Monte Carlo mixing for accurate estimation of the posterior distribution under both models, for the SDR application in Section \ref{app} and the simulation studies in Section \ref{sim}.

\subsection{Laplace Mechanism under Local DP}
\label{methods:Dwork}


As a comparison to our two Bayesian microdata synthesizers that use smoothing to encode disclosure protection, we include the alternative of adding noise to tabular products produced directly from the confidential survey data. Each product (e.g., cell means, cell counts, and corresponding standard errors) has a different amount of noise added from the Laplace distribution. See Supplementary Materials for a review of the Laplace Mechanism. We present sensitivity calculations for adding noise according to the Laplace distribution, which are based on the outcome variable and survey weights \emph{local} to the database $(\bm y_n, \bm X_n, \bm w_n)$. Since both the weights and outcomes are not required to be bounded, \emph{global} additive noise mechanisms do not exist (i.e., they would add noise of infinite or unbounded scale).
Let $\mathcal{S}_{f,g}$ represent the set of observations in field of study $f$ and gender $g$. Consistent with our use of partial synthesis for FBS and FBP, we assume that the unweighted sample sizes $n_{f,g}$ are not sensitive (i.e., publicly available) which is often the case for demographic surveys.

The local sensitivity $\Delta_{f,g}^c$ for count of field $f$ and gender $g$ (cell count) is:
\begin{equation}
\Delta_{f,g}^c = \max_{i \in \mathcal{S}_{f,g}}  w_i - \min_{i \in \mathcal{S}_{f,g}} w_i.
\label{eq:Delta1}
\end{equation}

The local sensitivity $\Delta_{f, g}^a$ for average salary of field $f$ and gender $g$ (cell average) is:
\begin{equation}
\Delta_{f, g}^a = \frac{\max_{i \in \mathcal{S}_{f,g}} w_i y_i - \min_{i \in \mathcal{S}_{f,g}} w_i y_i}{\sum_{i \in \mathcal{S}_{f,g}} w_i - (\max_{i \in \mathcal{S}_{f,g}} w_i - \min_{i \in \mathcal{S}_{f,g}} w_i)}.
\label{eq:Delta2}
\end{equation}

The marginal and total counts and averages are calculated in a similar fashion, once $\mathcal{S}_f, \mathcal{S}_g$, and $\mathcal{S}$ are defined accordingly. Finally, with calculated local sensitivity $\Delta_{f,g}^c$ for the count of field $f$ and gender $g$, the noise to be added to that cell count is sampled from the following Laplace distribution \citep{Dwork:2006:CNS:2180286.2180305}:
\begin{equation}
 \textrm{Laplace}(0, \Delta_{f, g}^c / \epsilon),
\end{equation}
where $\epsilon$ is the privacy budget. A similar process is used when adding noise to the average salary value of field $f$ and gender $g$ with calculated local sensitivity $\Delta_{f, g}^a$. For a given privacy budget $\epsilon$, the larger the local sensitivity, $\Delta_{f,g}^c$ or $\Delta_{f,g}^a$, the larger the scale for the added noise. It will often be the case that the observed values of $y_i$ and $w_i$ within each field $f$ and gender $g$ cell do not represent the full range of values in the corresponding cell population, so
we take the maximum of the cell-specific sensitivities and use that instead for the cell level Laplace noise:
$\Delta_{*}^{c} = \max_{f,g}\Delta_{f,g}^c$ and 
$\Delta_{*}^{a} = \max_{f,g}\Delta_{f,g}^a$.

We acknowledge that our implementation of the Laplace Mechanism adds noise to each cell, which will not ensure features such as the counts of male and female adding up to the count of both genders in a given field. Some modifications can be made to the query or the post-processing to enforce these consistency requirements \citep{li2010optimizing}. There are also works on post-processing optimization \citep{li2014DAWA}, which would benefit our synthetic data approaches as well once tables are created. For illustration in this paper, we do not pursue such modifications.

In addition to the counts and averages, we must also add noise to their corresponding variance estimates. We use a replication method \citep{Rao92} for a set of $R = 10$ replicates. We choose the random replication method of \cite{preston2009rescaled}, as opposed to the stratified jackknife or the balanced repeated replication (BRR), so that we can tune the number of replicates, $R$, directly.
 We add noise to each of these $10$ replicated point estimates, based on the sensitivity calculations for point estimates above to achieve a target $\epsilon_{vc} = 10 \epsilon_{rep}$ for a given cell, where ``$vc$" denotes ``variance cell".
 \begin{itemize}
     \item For each set of $R$ replicate weights $(w_i)_{r}$ we modify values for a subset for clusters. The method of \cite{preston2009rescaled} randomly selects half of the clusters in each strata, setting the weights of the others to $0$.
     Non-zero weights are then doubled such that weight totals are invariante across strata: $\sum_{i\in S}(w_i)_{r} = \sum_{i\in S} w_i = \hat{N}$.
     \item Estimate $R = 10$ sets $\hat{N}^{r}_{fg} = \sum_{i \in S} \mathbbm{1}^{i}_{fg} w^{r}_i $ and
$\hat{\mu}^{r}_{fg} = (\sum_{i \in S} \mathbbm{1}^{i}_{fg} w^{r}_i y_i)/\hat{N}^{r}_{fg}$.
     \item Estimate variance $\widehat{Var(\hat{\mu}_{fg})} = \frac{1}{R } \sum_{c \in S}(\hat{\mu}^{r}_{fg} - \hat{\mu}_{fg})^2$.
 \end{itemize}
 Then calculating between cluster variance $\widehat{Var(\hat{\mu}_{fg})}$ is a post-processing step performed on the set of $\hat{\mu}^{r}_{fg}$, and does not incur additional privacy loss.

Each individual's data is used in four table cells (one interior cell, plus the row and column margin, and the grand margin) to calculate two estimates (point and variance). There are two tables produced (counts and means). Assuming equal budget across each of these 8 pairs of estimates, then the total $\epsilon = 8 \epsilon_{pc} + 8 \epsilon_{vc}$, where ``$pc$" denotes ``point (estimate) cell". If we choose to assign equal budget to the point estimates and variance estimates, then setting a global budget of say $\epsilon = 8$, we use $\epsilon_{pc} = \epsilon/16 = 0.5$ for each cell point estimate, and $\epsilon_{rep} = \epsilon/ (16)(10) = 0.05$ for each of the 10 replicates, leading to $\epsilon_{vc} = 0.5$.


The FBS and FBP pseudo posterior and Laplace mechanisms all base their local privacy guarantees on the same observed database $(\bm y_n, \bm X_n, \bm w_n)$ and their local $\epsilon_{\bm y_n}$ levels are set to be equal for our analyses in the sequel to compare utility performances.  Only the pseudo posterior mechanisms, however, are equipped with an asymptotic global differential privacy guarantee.  The local DP guarantee of the Laplace Mechanism, by contrast, is based on the observed range for the outcome and weight values, which will vary between datasets.

\section{Application to the Survey of Doctorate Recipients}
\label{app}
In our real data application, we apply our two microdata synthesis approaches and the Laplace Mechanism to a sample of the SDR. 
Our sample comes from the public use file of 2017. The sample contains information on salary, field of study (8 levels), and gender (2 levels). For ease of illustration, we subset to the $n = 10,355$ respondents who are employed and from the 40-44 age group.

Our goal is to create survey tables of counts and average salary by field and gender along with corresponding standard error estimates, all with privacy protection. We fit our two microdata synthesis models presented in Sections \ref{methods:FBS} and \ref{methods:FBP} to generate synthetic survey microdata, from which we create survey tables containing point estimates and standard error estimates. As a comparison method, we add noise to the point estimates and standard error estimates using the Laplace distribution with local sensitivities for the Laplace Mechanism outlined in Section \ref{methods:Dwork}. Note that we scale the maximum Lipschitz bound in our two microdata synthesis approaches to express an equivalent value for the Lipschitz bounds, denoted as $\Delta_{\bm \alpha, (\bm y_n, \bm X_n, \bm w_n)}$. 
Therefore, we use $\epsilon_{\bm y_n} = 2 \Delta_{\bm \alpha, (\bm y_n, \bm X_n, \bm w_n)} \times  m$ as the total privacy budget for adding Laplace noise under the Laplace Mechanism comparison method, where $m$ denotes the number of replicate synthetic databases generated.  In this application, we use $m = 3$ based on the results of our simulation study that follows this application.

\subsection{A Note on Informative Sampling Designs}
\label{app:informative}

In general, data disseminators who are trying to create a private population estimator from samples using survey weights are faced with a challenge: the large variation in the survey design weights results in increased sensitivity of any additive noise privacy mechanism. Our microdata synthesis models, FBS and FBP, tackle this challenge by co-modeling the outcome variable and the survey weights. Therefore, under an informative sampling design, our methods would remove variation in weights unrelated to the outcome variable. The smoothed weights would be expected to produce improved utility, defined as preserving the sample-based tabular statistics, over additive noise mechanisms, such as the Laplace Mechanism that uses the raw sampling weights. 

However, our motivating SDR sample utilizes a nearly non-informative design (the correlation between the outcome variable salary and the survey design weights is about 0.09), and thus nearly representative of the population. We, therefore, expect to see: (a) The Laplace Mechanism performs well on counts, which are based on the survey weights only. The FBS and FBP perform similarly on counts compared to the Laplace Mechanism due to the limited value of weight smoothing for weights nearly uncorrelatd with the outcome; (b) FBS and FBP  produce private average salary values with higher utility than the Laplace Mechanism.  We see in the sequel that our methods perform dramatically better for utility preservation at equivalent privacy guarantees on average salary values. Both of our methods use a model to smooth the confidential data distribution as a means of encoding the formal privacy guarantee.  By contrast, the Laplace and related mechanisms add noise to the confidential data.   Under our use of a modeling framework, we are able to selectively downweight record-indexed likelihood contributions to be more precise in encoding privacy, which better preserves distributional characteristics in the synthetic data.

If the sampling design is informative, which is the more common setup that we explore in our simulation studies of Section \ref{sim}, our FBS and FBP are expected to outperform the Laplace Mechanism on both counts and average salary values, due to their co-modeling of salary and weight that produces weight smoothing for more efficient estimators.

\subsection{Model Fit Performances}
\label{app:fit}

\begin{figure}[t]
  \centering
   \includegraphics[width=0.8\textwidth]{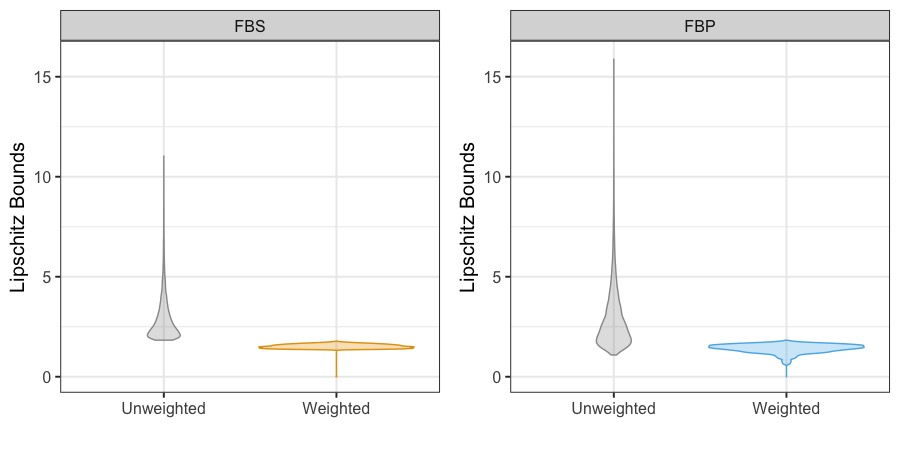}
      \caption{Distributions of record-level Lipschitz bounds of the non-private unweighted and the private weighted of FBS (left) and FBP (right) in the SDR application.}
      \label{fig:app_Lbounds_app1app2}
\end{figure}

We next investigate the model fit performances of FBS and FBP. Each plot panel in Figure \ref{fig:app_Lbounds_app1app2} plots distributions of the record-level Lipschitz bound $\Delta_{\bm \alpha, (y_i, \bm X_i, w_i)}$ for each of FBS and FBP under the $\bm{\alpha}-$weighted pseudo posterior mechanism.  Each plot panel also includes a plot for the unweighted posterior mechanism, achieved by setting all of the $\alpha_{i} = 1$ in each of FBS and FBP for comparison. The maximum value of each violin plot of the weighted on the y-axis corresponds to the overall Lipschitz bound, $\Delta_{\bm \alpha, (\bm y_n, \bm X_n, \bm w_n)}$ of each approach. As Figure \ref{fig:app_Lbounds_app1app2} shows, our (weighted) FBS and FBP have equivalent maximum Lipschitz bound, about 1.8, indicating that each approach provides an ($\epsilon_{\bm y_n} = 2 \times 1.8 \times 3 = 10.8$) asymptotic differential privacy guarantee with $m = 3$ simulated synthetic datasets. Note that the overall Lipschitz bounds of the two approaches under the unweighted, i.e., without $\bm \alpha$ weighting, are 11.0 and 15.9 respectively (the maximum value of each unweighted violin plot on the y-axis) 
leading to $\epsilon_{\bm y_n}$ values of 66.0 and 95.4, respectively. 
Our $\bm \alpha-$weighted approaches produce substantially lower overall Lipschitz bounds, indicating the ability of vector weights $\bm \alpha$ to control $(\epsilon_{\bm y_n} = 2 \Delta_{\bm \alpha, (\bm y_n, \bm X_n, \bm w_n)} \times m)-$DP privacy guarantee with $m$ synthetic datasets. 

Figure \ref{fig:app_Lbounds_app1app2} further reveals that the (weighted) Lipschitz distribution for FBS produces more records expressing relatively high values of $\Delta_{\bm \alpha, (y_i, \bm X_i, w_i)}$ that are concentrated around the overall Lipschitz, $\Delta_{\bm \alpha, (\bm y_n, \bm X_n, \bm w_n)}$ for the entire database as compared to FBP. Although only the overall Lipschitz bound (i.e., the maximum) controls the overall privacy guarantee, we observe that FBS avoids overly downweighting records' likelihood contributions through $\bm \alpha$, which will result in higher utility as we will see in Section \ref{app:utility}, making it a more efficient synthesizer \citep[See also][]{SavitskyHuWilliams2020reweight}.

It bears noting that for real data applications, such as Google's 2020 COVID-19 Mobility Reports and LinkedIn's Audience Engagement API, the privacy budgets being used ranged from 8.6 to 79.22 for monthly queries \citep{BowenGarfinkel2021Notices}. The SDR is conducted every other year, and we believe a privacy budget of 10.8 is within an acceptance range in real data applications given current practices. Moreover, we note that releasing $\Delta_{\alpha, \bm{y}_n}$, such as in Figure \ref{fig:app_Lbounds_app1app2}, is equivalent to releasing $\epsilon_{\bm{y}_n}$ which contracts on the global $\epsilon$ for $n$ sufficiently large (more than a couple of hundred units). In other words, there is no additional leakage by releasing $\Delta_{\alpha, \bm{y}_n}$.

\subsection{Utility Evaluation}
\label{app:utility}

With both Bayesian methods satisfy equivalent privacy guarantee, we compare their utility results, the point estimates and standard error estimates of the resulting privacy protected tables. Our methods are also compared to those of noise-added point and standard error estimates from the Laplace Mechanism satisfying equivalent privacy guarantee of $\epsilon_{\bm y_n} = 10.8$. 

We recall that our goal is to create private survey tables of counts of observations and average salary values by field and gender that contain point estimates and standard error estimates. Given the fact that we have 8 levels in field and 2 levels in gender, we have 27 cells for counts: $8 \times 2 = 16$ cells for a field and gender combination, $8$ cells for both genders for $8$ field, $2$ cells for all fields for $2$ genders, and lastly $1$ cell for all fields and both genders. We also have 27 cells for average salary values with the same breakdown.

\begin{figure}[t]
  \centering
   \includegraphics[width=0.8\textwidth]{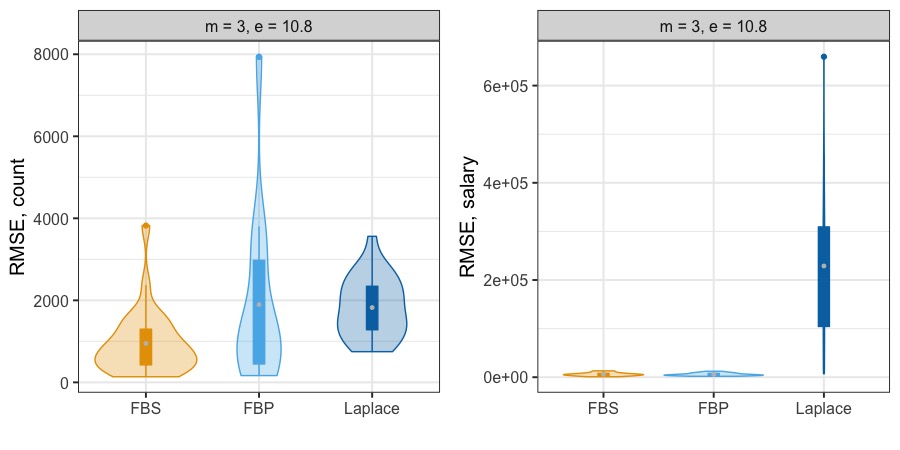}
   \caption{RMSE values of {\bf{counts}} (left) and {\bf{average salary values}} (right) of the three methods, FBS, FBP, and Laplace, applied to the SDR sample. Each violin plot represents a distribution of RMSE values over 27 cells. Results are based on $m = 3$ synthetic datasets by FBS and FBP, achieving $\epsilon_{\bm y_n} = 10.8$ for all three methods.}
      \label{fig:app_RMSE_count_salary_L1p8_m3_app1app2Laplace}
\end{figure}

For each cell, we obtain a point estimate and a standard error estimate for each of the three methods, FBS, FBP, and the Laplace Mechanism. We utilize the point estimate constructed from the underlying confidential sample to calculate a root mean square error (RMSE), defined as the square root of the total of the squared bias and the variance for each of the three methods. The smaller the RMSE value, the higher the utility. We compute an RMSE value for each of the 27 cells for each method, and Figure \ref{fig:app_RMSE_count_salary_L1p8_m3_app1app2Laplace} shows the distributions of RMSE values under all three methods, for counts (left) and average salary values (right). The cell-level comparison tables are available in Tables \ref{tab:sdr:counts_RMSE} and \ref{tab:sdr:salary_RMSE} for further reading. Our discussion of the results mainly focuses on Figure \ref{fig:app_RMSE_count_salary_L1p8_m3_app1app2Laplace}.

\begin{table}[t]
\centering
\begin{tabular}{ l | r |  r r  r }
\hline
  & Sample & Laplace&  FBS  & FBP \\ \hline
 All fields & 125199 (1383) &  125323 (2281)& 125199 (234) &  125199 (165) \\
 Male & 76610 (1305)  & 76917 (964)& 76583 (876) & 68696 (638) \\
 Female & 48589  (930) & 48510 (744) & 48616 (753) & 56503 (610) \\ \hline
 Field 1 total & 34185 (725)  & 34255 (2732) &  30370 (247) & 37505 (567) \\ 
 Male  & 18545 (652) & 18526 (2402) & 17089 (441) & 18528 (441) \\ 
 Female & 15640 (533) & 15606 (1981) & 13281 (250) & 18977 (440) \\  \hline
 Field 2 total & 5555 (280) & 5859 (2401) & 6513 (194) & 4706 (246) \\ 
 Male  & 4495 (295)  & 4757 (1824) & 4848 (205) & 3324 (209) \\ 
 Female  & 1059 (134)  & 1115  (757) & 1665 (155) &  1382 (134) \\ \hline
 Field 3 total & 6085 (356)  & 6160 (2128) & 5934 (256) & 4673 (235) \\ 
 Male & 4351 (331)  & 4435 (1010) & 4362 (219) & 3197 (197) \\ 
Female  & 1734 (237) & 1736 (2142) & 1572 (154) & 1476 (132) \\  \hline
 Field 4 total & 18539 (529) & 18504 (3561) & 17625 (195) & 21457 (469) \\ 
 Male  & 13714 (527)  & 13728  (816) & 12310 (283)  & 13653 (391)  \\ 
 Female  & 4825 (277) & 4931 (1486) & 5315 (185) & 7804 (299) \\  \hline
 Field 5 total & 13238 (415)  & 13329 (2726) & 14779 (475) & 11730 (358) \\ 
 Male  & 4312 (339) & 4290 (1440) &  5074 (290) & 3439 (204)  \\ 
 Female  & 8926 (369)  & 8983 (2165) & 9705 (400) & 8291 (304) \\ \hline
 Field 6 total & 15810 (448) & 15784 (1691) & 16991 (371) & 15789 (405)  \\ 
 Male  & 8328 (441)  & 8542 (1267) & 8664 (361) & 7045 (284) \\ 
 Female  & 7483 (313) & 7431 (1533) &  8327 (260) & 8744 (308) \\ \hline
 Field 7 total & 26977 (688) & 26940 (2872) & 28600 (133) & 23376 (506) \\ 
 Male & 20822 (673)  & 21104 (1746) & 22131 (353) & 17044 (443) \\ 
 Female  & 6156 (400)  & 6110 (1372) & 6468 (272) & 6332 (280) \\ \hline
 Field 8 total & 4810 (238) & 4686 (1274) & 4388 (121) & 5963 (264) \\ 
 Male & 2044 (206) & 1899 (1127)& 2105 (123) & 2466 (174) \\ 
 Female  & 2766 (192) & 2819 (2650) & 2283 (118) & 3497 (203) \\ \hline
\end{tabular}
\caption{Point and standard error estimates of counts: survey weighted expansion estimate of the population values based on the observed sample  (Sample), noise added from the Laplace distribution to the observed sample (Laplace), and our microdata synthesis approaches of fully Bayes for observed sample (FBS) and Fully Bayes for Population (FBP).}
\label{tab:sdr:counts_RMSE}
\end{table}

\begin{table}[t]
\centering
\begin{tabular}{ l | r |  r r r }
\hline
  & Sample & Laplace&  FBS & FBP \\ \hline
 All fields & 108707 (1050)  & 109513 (11190)& 107421 (780) & 107031 (964)\\   
 Male & 115785 (1491)  &  115664  (7352) & 113623 (1013) & 112206 (1445)  \\
 Female & 97548 (1309)  & 99593 (18564) & 97660  (959) & 100840 (1206) \\ \hline
 Field 1 total & 107609 (2133)  & 107017  (42677) & 97568  (954) & 97392 (1481)  \\ 
 Male & 115172  (3299)   & 101900 (440042) & 102378 (1419) & 102984 (1575) \\ 
 Female  & 98641  (2478)  & 102608 (223808) & 91366 (1368) & 92099 (1998)  \\ \hline
 Field 2 total & 132192 (4874) & 135918 (174237) & 135373 (4204) & 136571 (4271) \\ 
 Male & 135522 (5648)  & 163687 (234994) & 137929 (5157) & 140664 (5544 \\ 
 Female  & 118061  (9376) & 134395 (218548) & 128013 (7900) & 126984 (6455)  \\ \hline
 Field 3 total & 104992 (4885)   & 105217  (55443) &  98907 (2758) & 101444 (2899) \\ 
 Male & 105941  (5464) & 99976 (354606) & 102080 (3399) & 104261 (3941) \\ 
 Female  & 102610 (10184)   & 130265 (342208) & 90094 (3767) & 95494 (4166) \\ \hline
 Field 4 total & 110323 (2530)  & 107428 (106783) & 107985 (1860) & 104866 (1319) \\ 
 Male & 111671  (3026) & 117144 (291832) & 112101 (2077)  & 109198 (1712) \\ 
 Female  & 106513  (4575) & 100038 (287057) & 98456 (2695) & 97510 (2498) \\ \hline
 Field 5 total &  94520 (2552)  & 89890 (175891) &  98409 (1791)  & 97013 (1736) \\ 
 Male & 104400  (5966)   & 92574 (326365) &  105570 (3429) & 103183 (3696) \\ 
 Female  & 89748  (2408) & 91096 (659405) &  94654 (2002) & 94515 (1828) \\ \hline
 Field 6 total & 95754 (3216)   & 101597 (119309) & 93209 (2202) & 92206 (1435)  \\ 
 Male & 104427  (5505)  & 58174 (265991) & 97993 (2951) & 98955 (2227) \\ 
 Female  & 86103  (2768) & 82758 (199529) & 88208 (2140) & 86964 (1721) \\ \hline
 Field 7 total & 120444 (2315)   & 118149 (101962) & 125963 (2302) & 125303 (2030) \\ 
 Male & 123570  (2766) & 117140 (459755) & 127912 (2577) & 128236 (2020) \\ 
 Female  & 109867  (3748)  & 129382 (402416) & 119289 (3379) & 117615 (4144) \\ \hline
 Field 8 total & 103636 (4147) & 104684  (99206) & 107302 (3945) & 101453 (2508) \\
 Male & 117473  (7344) & 115349 (274175) & 114344 (5049)  & 110167 (4458) \\ 
 Female  & 93411  (4236) & 89180 (276217) & 100778 (4668) & 95536 (3055)  \\ \hline
\end{tabular}
\caption{Point and standard error estimates of average salary values: survey weighted expansion estimate of the population values based on the observed sample  (Sample), noise added from the Laplace distribution to the observed sample (Laplace), and our microdata synthesis approaches of fully Bayes for observed sample (FBS) and Fully Bayes for Population (FBP).}
\label{tab:sdr:salary_RMSE}
\end{table}

For the counts, FBS slightly outperforms the Laplace Mechanism, even under a relatively non-informative sampling design as the SDR sample. FBP, by contrast, expresses more variation in the RMSE result due to the complex construction of the exact likelihood in the observed sample.  This method is, perhaps, not designed to use the smoothed weights since the outcome variable is synthesized as unbiased with respect to the population. On the average salary values, FBS and FBP dramatically outperform the Laplace Mechanism, due to their co-modeling of the salary variable and the survey weights. Between the two microdata synthesis methods, FBP produces an even more contracted RMSE distribution. This is because FBP incorporates correction for population bias of the salary variable in the model. 

In addition, our implementation of the Laplace Mechanism does not by default maintain certain key features, such as the counts of male and female adding up to the count of both genders in a given field, which are naturally maintained by the microdata synthesis approaches, as evident in Tables \ref{tab:sdr:counts_RMSE} and \ref{tab:sdr:salary_RMSE}. Improvements can be made to enforce these consistency requirements, such as methods proposed by \citet{li2010optimizing}, which are not pursued in this paper.

Overall, our FBS and FBP create sample-based tabular statistics with higher utility than the Laplace Mechanism for the SDR application. Moreover, FBS and FBP produce private synthetic microdata that is readily available for public release for additional analyses by users, whereas the Laplace Mechanism only produces private tabular products. Between the two, FBS performs much better on counts while FBP slightly performs better on average salary values. We also note that FBS has higher modeling flexibility and is more straightforward to implement than FBP. 

As discussed in Section \ref{app:informative}, when working with a more informative sampling design, we anticipate our FBS and FBP to outperform the Laplace Mechanism on both counts and average salary values. By co-modeling the salary variable and the survey weights, FBS and FBP will remove the variation in survey weights through smoothing, which in turn creates smaller standard error estimates for the cells. The Laplace Mechanism, on the other hand, will suffer greatly from the variation of survey weights due to increased sensitivity values. To fully explore such scenarios, next we conduct a series of simulation studies where we utilize an informative sampling design that is more typical of those administrated by government statistical agencies. 

\section{Simulation Studies}
\label{sim}

In our simulation studies to follow we take a sample from a simulated population under an informative sampling design. We perform estimation on the observed sample using the FBS and FBP approaches for synthesis and create survey tables from the resulting synthetic microdata samples. As in the SDR application, we scale the two approaches at equivalent levels of privacy guarantee to compare their utility performances. We also add noise under the Laplace Mechanism with the equivalent level of privacy guarantee to the table formed by the simulated, confidential sample, and include it for utility comparison. 

We introduce the simulation design with informative sampling, investigate the model fit performances, and conduct utility evaluations in Sections \ref{sim:design} to \ref{sim:utility}. We evaluate the utility-risk trade-off of our proposed methods in Section \ref{sim:tradeoff}.  Moreover, we discuss findings from comparing FBS and FBP under repeated sampling, where the results are included in the Supplementary Materials for further reading.

\subsection{Simulation Design}
\label{sim:design}

We design our simulation study based on the public use file of the 2017 SDR. We simulate a population of $N = 100,000$ units containing unit-level information on salary, field, and gender. In our simulated population, the field and gender percentages follow those in the public use file. Given simulated field and gender, each unit's salary value $y_i$ is simulated from a lognormal distribution with a field and gender specific mean (obtained from the public use file) and a fixed scale of 0.4. We simulate inclusion probability $\pi_i$ for unit $i$ by generating additive $\textrm{noise}_i$ from a normal distribution with 0 mean and the same fixed scale of 0.4 in the following construction,
\begin{equation}
\label{eq:sw}
\log (\pi_i) = \log (y_i) + \textrm{noise}_i.
\end{equation}
We then obtain survey weight $w_i \propto 1 / \pi_i$\footnote{ The actual marginal inclusion probabilities $\pi^{*}_i = 1-(1-  \pi_i / \sum_{i \in h}\pi_i)^{n} \approx n* \pi_i / \sum_{i \in h}\pi_i \propto \pi_i$ and the sample is selected with replacement. This is close to sampling without replacement when using small sampling fractions. The $\pi^{*}_i$'s are calculated with the \texttt{inclusionprobabilities()} function from the \texttt{sampling} R package, which computes the first-order inclusion probabilities for a probability proportional-to-size sampling design \citep{sampling}.}. Less noise corresponds to a more informative sampling design, whereas more noise corresponds to a weaker relationship between the outcome and the selection probability. We choose a moderate level of noise corresponding to a moderately informative design, resulting in a $-0.57$ correlation between the salary and the survey weights in the population.

Next, we take a stratified probability proportional to size (PPS) sample of $n = 1000$ units, where fields are used as strata (and $\pi_i$ is used as the size variable in Equation~(\ref{eq:sw})). We denote $\bm y_n$ as the outcome variable salary, $\bm X_n$ as the field and gender variables, and $\bm w_n$ as the sampling weights of the realized / observed sample. The correlation between $\bm y_n$ and $\bm w_n$ is -0.58 in the sample.

\subsection{Model Fit Performances}
\label{sim:fits}

\begin{figure}[t]
  \centering
   \includegraphics[width=0.8\textwidth]{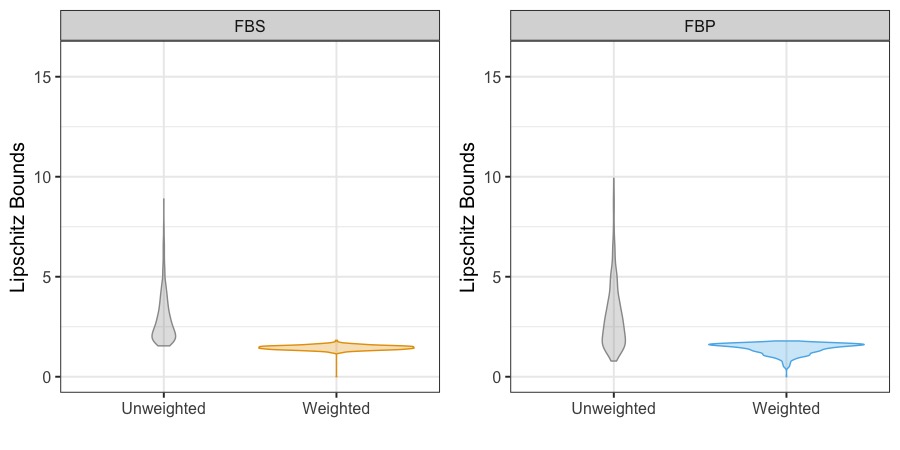}
      \caption{Distributions  of  record-level  Lipschitz  bounds  of  the  non-private  unweighted  and the private weighted of FBS (left) and FBP (right) in the simulation.}
      \label{fig:sim_Lbounds}
\end{figure}

To investigate the model fit performances of FBS and FBP pseudo posterior mechanisms on the simulated sample, we examine the distributions of record-level Lipschitz bounds $\Delta_{{\bm \alpha}, (y_i, {\bm X_i}, w_i)}$ of FBS and FBP in Figure \ref{fig:sim_Lbounds}. As before, their corresponding unweighted posterior mechanism Lipschitz bounds distributions are included for comparison. Our (weighted) FBS and FBP have an equivalent maximum Lipschitz bound of about 1.8, which means that both approaches provide an $(\epsilon_{\bm y_n} = 2 \times 1.8 \times 3 = 10.8)$ asymptotic differential privacy guarantee with $m = 3$ simulated synthetic datasets. The maximum Lipschitz bounds of the two unweighted are 8.89 and 9.92, respectively. As in the SDR application, our $\bm \alpha-$weighted FBS and FBP pseudo posterior mechanisms produce lower overall Lipschitz bounds and provide an $(\epsilon_{\bm y_n} = 10.8)$ privacy guarantee. Moreover, FBS downweights less of records' likelihood contributions through $\bm \alpha$, which is expected to result in higher utility.

\begin{figure}[t]
  \centering
   \includegraphics[width=0.8\textwidth]{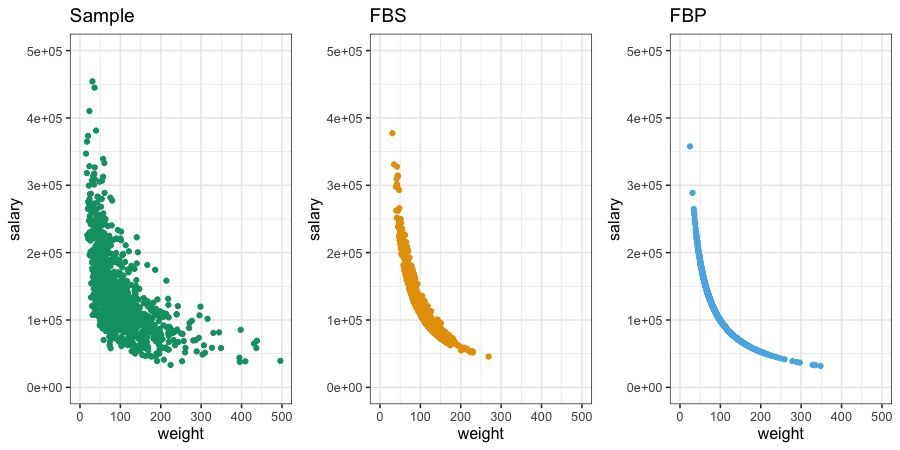}
      \caption{Comparison of the salary and weight bivariate distributions of confidential salary and weights in the sample (green and left), synthetic salary and smoothed weights from FBS (yellow and middle), and synthetic salary and smoothed weights from FBP (blue and right).}
      \label{fig:sim_salary_weight_L1p8}
\end{figure}


Figure \ref{fig:sim_salary_weight_L1p8} displays the weight smoothing effects of our FBS and FBP in relation to the salary variable. Compared to the bivariate distribution of confidential weights and confidential salary in the sample, FBS and FBP produce smoothed weights that show much less variation. Between the two methods, FBP creates slightly smoother estimation of the weights than FBS. Even though we do not use the smoothed weights directly in FBP to construct average salary values (as we do in FBS), the smoothed weights in FBP allows a more precise estimation of the exact posterior distribution for the sample. Moreover, FBP incorporates correction for population bias of the salary variable by design. We therefore expect to see higher utility of average salary values produced by FBP. Additional results comparing smoothed weights to synthesized weights in our FBS and FBP are available in the Supplementary Materials.

\subsection{Utility Evaluation}
\label{sim:utility}

We proceed to compare the their utility results, defined as the point estimates and standard estimates of the resulting private tables achieved by the three methods (FBS, FBP, Laplace) under an equivalent privacy guarantee.

\begin{figure}[t]
  \centering
   \includegraphics[width=0.8\textwidth]{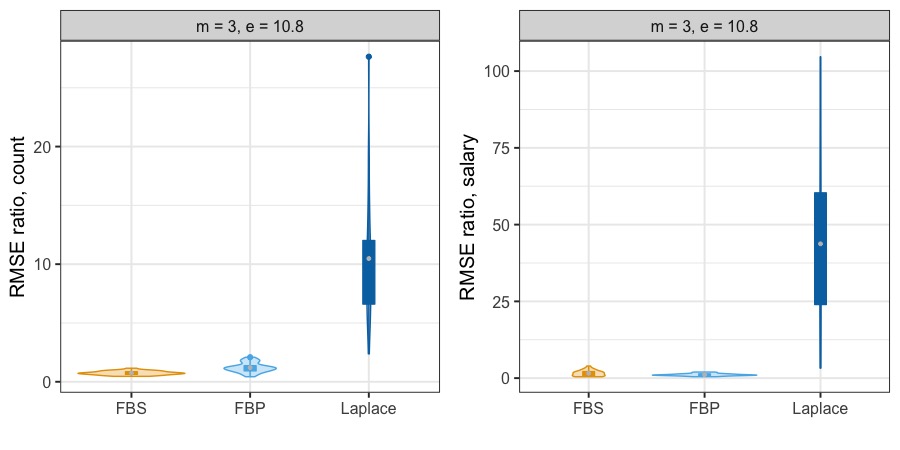}
      \caption{RMSE ratios of {\bf{counts}} (left) and {\bf{average salary values}} (right) of the three methods, FBS, FBP, and Laplace, applied to the selected sample. Each violin plot represents a distribution of RMSE ratios over 27 cells. Results are based on $m = 3$ synthetic datasets by FBS and FBP, achieving $\epsilon_{\bm y_n} = 10.8$ for all three methods.}
      \label{fig:sim_RMSEratio_Count_Salary_L1p8_m3}
\end{figure}

Unlike the SDR application where we do not know the population truth, in our simulation studies we \emph{know} the population value of counts and average salary values of all cells of interest, which involve 27 cells for counts and another 27 cells for average salary values. Therefore, we calculate the RMSE value of each cell of each of the three methods compared to the cell's population value. We produce the corresponding RMSE value based on the confidential sample, and create an RMSE ratio, defined to be the cell-specific RMSE value of our methods over that of the sample. The smaller the RMSE ratio, the higher the utility. We present the distributions of 27 RMSE ratios for counts (left) and average salary values (right) in Figure \ref{fig:sim_RMSEratio_Count_Salary_L1p8_m3}. 

On the counts, FBS and FBP clearly outperform the Laplace Mechanism, with FBS producing the smallest RMSE ratios and therefore the highest utility. The superior performance of FBS and FBP lies in their co-modeling of the outcome salary and the survey weights. While the specific synthesizing model is different between them, both methods take the advantage of co-modeling the weights and outcome, which lead to weight smoothing and result in more stable estimates of domain counts which are tabulated from the weights \citep[See for example][]{beaumont2008new}. This is especially true when an informative sampling design is employed, as we do here in our simulation design. Between the two methods, FBS performs better. As shown in the Supplementary Materials under repeated sampling, FBP overcovers by producing longer confidence intervals, which explain its higher RMSE ratio values. 

For average salary values, FBS and FBP perform dramatically better than the Laplace Mechanism. These results once again illustrate the main advantage of co-modeling of our FBS and FBP over the Laplace Mechanism: to fully utilize any information in the sampling weights to improve estimation of the outcome variable(s). Between the two methods, FBP shows a more contracted RMSE ratio distribution, indicting higher utility than FBS. The advantage of FBP lies in its incorporation of population bias by design and its enhanced weight smoothing in relation to the salary (as shown in Figure \ref{fig:sim_salary_weight_L1p8} and discussed in Section \ref{sim:fits}).

By contrast, when implementing the Laplace Mechanism, given a fixed privacy budget $\epsilon_{\bm y_n}$, the amount of noise to be added solely depends on the cell sensitivity: the scale of Laplace noise is proportional to the cell sensitivity, so that larger cell sensitivity results in larger noise to be added \citep{Dwork:2006:CNS:2180286.2180305}. When units are accompanied with sampling weights, and if the sampling weight distribution has a large variability, as is often the case in practice, the cell sensitivity of counts can be large. Moreover, as with the SDR application in Section \ref{app:utility}, the Laplace Mechanism does not maintain certain key features, such as the counts of male and female adding up to the count of both genders in a given field, which are naturally maintained by FBS and FBP (results omitted for brevity).


In summary, our microdata synthesis approaches outperform the Laplace Mechanism in utility preservation of point estimates and standard error estimates, illustrated by the RMSE ratio metric. Between the two methods, FBS produces counts with higher utility while FBP produces average salary with higher utility. Overall both methods perform reasonably well across the two sets of tabular statistics. We next illustrate how to tune the utility-risk trade-off of FBS and FBP, by investigating the effects of $m$, the number of simulated synthetic datasets, and the overall Lipschitz bounds.

\subsection{Tuning the Utility-Risk Trade-off}
\label{sim:tradeoff}

Both our FBS and FBP approaches provide $(\epsilon_{\bm y_n} = 2 \times \Delta_{{\bm \alpha}, (\bm y_n, \bm X_n, \bm w_n)} \times m)-$DP privacy guarantee with $m$ synthetic datasets. Our results so far are all based on simulating $m = 3$ synthetic datasets, which given $\Delta_{{\bm \alpha}, (\bm y_n, \bm X_n, \bm w_n)} = 1.8$, provide an $(\epsilon_{\bm y_n} = 2 \times 1.8 \times 3 = 10.8)$ privacy guarantee. In this section, we investigate the utility-risk trade-off of the FBS and FBP approaches from two angles: the effects of various choice for $m$, the number of simulated synthetic datasets, and the scaling and shifting our weights, $\bm{\alpha}$ to shift $\Delta_{{\bm \alpha}, (\bm y_n, \bm X_n, \bm w_n)}$, both of which are positively linear with the overall privacy guarantee, according to the $(\epsilon_{\bm y_n} = 2 \times \Delta_{{\bm \alpha}, (\bm y_n, \bm X_n, \bm w_n)} \times m)$ result.

\begin{figure}[t]
  \centering
   \includegraphics[width=0.8\textwidth]{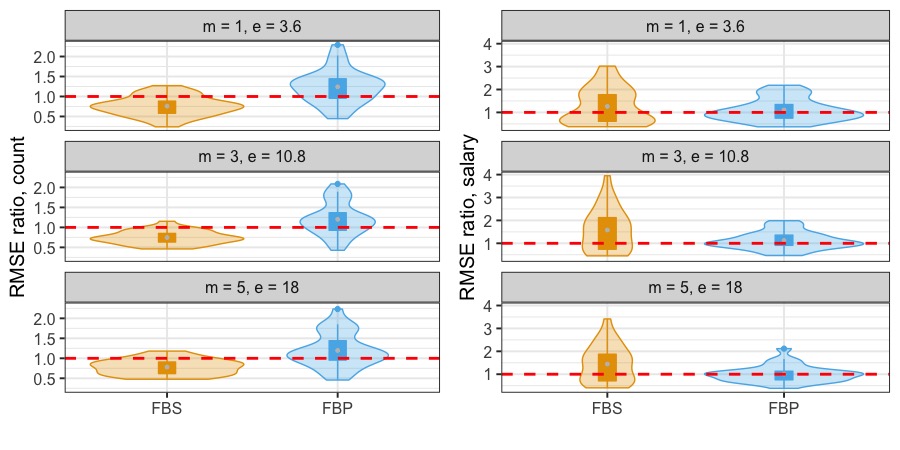}
      \caption{RMSE ratios of {\bf{counts}} (left) and {\bf{average salary values}} (right) of FBS and FBP, applied to the selected sample. A red dashed line at RMSE ratio = 1 is included for reference. Each violin plot represents a distribution of RMSE ratios over 27 cells. Results are based on $m = \{1, 3, 5\}$ synthetic datasets by FBS and FBP, achieving $\epsilon_{\bm y_n} = \{3.6, 10.8, 18\}$ for both methods.}
      \label{fig:sim_RMSEratio_Count_Salary_L1p8_3m_app1app2}
\end{figure}

The first set of investigation on the effects of $m$ requires no additional model fits of FBS and FBP. Given the current model fits which result in equivalent overall Lipschitz bounds of $\Delta_{{\bm \alpha}, (\bm y_n, \bm X_n, \bm w_n)} = 1.8$, we experiment with different values of $m$ and evaluate its impact on the privacy budget $(\epsilon_{\bm y_n} = 2 \times 1.8 \times m)$ and its RMSE ratio utility. We choose to experiment with the set of $m = \{1, 3, 5\}$, which are associated with $\epsilon_{\bm y_n} = \{3.6, 10.8, 18\}$ for both methods. Figure \ref{fig:sim_RMSEratio_Count_Salary_L1p8_3m_app1app2} displays the RMSE ratio utility results of counts (left) and average salary values (right), where each panel corresponds to a specific $(m, \epsilon_{\bm y_n})$ combination, over 27 cells. The middle panel in each side corresponds to $(m = 3, \epsilon_{\bm y_n} = 10.8)$, the results previously shown in Figure \ref{fig:sim_RMSEratio_Count_Salary_L1p8_m3}. These enlarged panels without the Laplace Mechanism clearly demonstrate that FBS performs better on counts while FBP performs better on average salary values. Moreover, in many cases FBS achieves smaller-than-1 RMSE ratios for counts and FBP produces smaller-than-1 RMSE ratios for average salary values, suggesting that the private FBS and FBP tabular data produce more efficient estimators than the non-private, confidential sample.

The impact of $m$ on the RMSE ratio utility metric is in accordance to the expected utility-risk trade-off for the most part: As $m$ increases from 1 to 3 and then to 5, the RMSE ratio distributions of both methods become slightly more contracted and overall smaller in for the counts, indicating slightly improved utility at the price of higher privacy budget (i.e., higher risks). This is true for FBP on average salary values, while increasing $m$ has little impact on FBS results on average salary values. Given the small utility improvement on estimated salary totals for the cells, increasing $m$ from 3 to 5, does not sufficiently improve utility to justify the required amount of added privacy budget (from 10.8 to 18 in this case). Therefore the $(m = 3, \epsilon_{\bm y_n} = 10.8)$ setup is ideal for a utility-risk trade-off balance {\color{blue}{in our simulation setting}}.

\begin{figure}[t]
  \centering
   \includegraphics[width=0.8\textwidth]{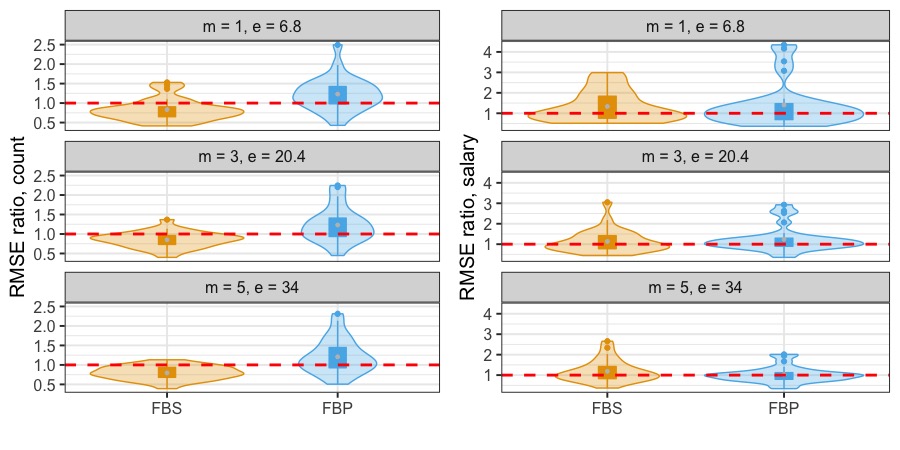}
      \caption{RMSE ratios of {\bf{counts}} (left) and {\bf{average salary values}} (right) of FBS and FBP, applied to the selected sample. A red dashed line at RMSE ratio = 1 is included for reference. Each violin plot represents a distribution of RMSE ratios over 27 cells. Results are based on $m = \{1, 3, 5\}$ synthetic datasets by FBS and FBP, achieving $\epsilon_{\bm y_n} = \{6.8, 20.4, 34\}$ for both methods.}
      \label{fig:sim_RMSEratio_Count_Salary_L3p4_3m_app1app2}
\end{figure}

For the second investigation on the effects of overall Lipschitz bound $\Delta_{{\bm \alpha}, (\bm y_n, \bm X_n, \bm w_n)}$, we refit the two methods on the selected sample and perform less downweighting overall to achieve a higher overall Lipscthiz bound, $\Delta_{{\bm \alpha}, (\bm y_n, \bm X_n, \bm w_n)} = 3.4$, for both methods. The refits require tuning the scaling and shifting constants, $(c_1, c_2)$ in creating new privacy protection weights $\alpha_i^* = c_1 \times \alpha_i + c_2$ to achieve a targeted overall Lipschitz bound $\Delta_{{\bm \alpha}, (\bm y_n, \bm X_n, \bm w_n)}$ \citep{SavitskyWilliamsHu2020ppm}. As before, we experiment with the set of $m = \{1, 3, 5\}$, which are associated with $\epsilon_{\bm y_n} = \{6.8, 20.4, 34\}$ for both methods. RMSE ratio results are shown in Figure \ref{fig:sim_RMSEratio_Count_Salary_L3p4_3m_app1app2}.

Although the overall Lipschitz bound $\Delta_{{\bm \alpha}, (\bm y_n, \bm X_n, \bm w_n)}$ increases from 1.8 to 3.4, which results in higher privacy budgets for a given $m$, we do not see much utility improvement for FBP, and in some cases the utility slightly deteriorates (note that the y-axis scale in Figure \ref{fig:sim_RMSEratio_Count_Salary_L3p4_3m_app1app2} left for counts goes up to 2.5 while that in Figure \ref{fig:sim_RMSEratio_Count_Salary_L1p8_3m_app1app2} goes up to 2). By contrast, the utility of FBS, especially on average salary values, shows notable improvement as $\Delta_{{\bm \alpha}, (\bm y_n, \bm X_n, \bm w_n)}$ increases from 1.8 to 3.4. It is then a decision for the data disseminators to strike a balance of the utility-risk trade-off when determining the ideal combination of $(m, \epsilon_{\bm y_n})$ given their dissemination goals and priority.

In addition, we compare the performances of FBS and FBP under repeated sampling through a Monte Carlo simulation study and the results are included in the Supplementary Materials for further reading. The repeated sampling results reinforce our findings based on one replicate of simulated sample, that FBS performs better on the counts while FBP performs better on average salary values. Moreover, our simulation setting benefits the most from the $(m = 3, \epsilon = 10.8)$ combination for the utility-risk trade-off. Note that here the database $\bm{y}_n$ is dropped from $\epsilon_{\bm{y}_n}$ in this $(m = 3, \epsilon = 10.8)$ notation because we are achieving the global $\epsilon$ that is shared amongst different $\bm{y}_n$ samples taken from the population.

In summary, based on the SDR application results and the simulation results, our FBS and FBP dramatically outperform the Laplace Mechanism in our setting. Combined with the modeling flexibility FBS, we conclude that FBS is the preferred microdata synthesis approach. In particular, FBS is very straightforward to implement and accommodates any desired model for the observed sample. Our experiments on $m$ and $\Delta_{\bm \alpha, (\bm y_n, \bm X_n, \bm w_n)}$ for tuning the utility-risk trade-off in our simulation prefers the $(m = 3, \epsilon = 10.8)$ combination. Data disseminators are encouraged to tune $m$ and $\Delta_{\bm \alpha, (\bm y_n, \bm X_n, \bm w_n)}$ and make decisions according to their dissemination goals and priority.

\section{Concluding Remarks}
\label{conclusion}

We address the issue of formal privacy for data collected from an informative sampling design. There are three major challenges for inducing formal privacy protection into survey data: (a) The correlation between weights and outcome variable(s) affects data utility since the distribution in the sample is different from the underlying population. It also makes privacy protection more complex because the sampling weights also need to be protected. (b) Under skewed and possibly unbounded outcomes (weights and salary) typically present for survey data, traditional additive noise mechanisms perform poorly and global privacy guarantees are typically not available due to unbounded variables and sampling weights that produce unbounded sensitivities. (c) Both standard errors and point estimates must be produced and released in a private manner. Standard error calculations from complex survey data are non-standard.

We develop and apply two modeling approaches to perform partial data synthesis, both of which co-model the outcome variable(s) and the survey weights, jointly: FBS performs modeling under the distribution for the observed sample, while FBP directly incorporates correction for population bias of the outcome variable(s).  FBS and FBP are specific implementations of our pseudo posterior mechanism that comes equipped with an asymptotic differential privacy guarantee.

Our SDR application with a nearly non-informative sampling design shows that FBS and FBP perform almost equally well on tabular cell counts and much better on average salary values, compared to the Laplace Mechanism. When the sampling design is informative, which is typically the case in practice, our simulation studies demonstrate the notably superior performances of FBS and FBP on both counts and average salary values as compared to the Laplace Mechanism. Their advantage over the Laplace Mechanism derives from two facets: (a) The targeted downweighting of only high-risk records under the pseudo posterior mechanisms used for both FBS and FBP limits the degree of distortion induced into the confidential data; (b) The joint modeling of the outcome variable and sampling weights produces more efficient smoothed weights by removing variation in the weights unrelated to the outcome.

Between the two methods, FBS consistently shows higher utility on tabular cell counts while FBP shows slightly better utility performance on average salary values. The Monte Carlo simulation under repeated sampling available in the Supplementary Materials favors FBS, overall. Given the modeling flexibility of FBS and its relative ease-of-implementation for any synthesizing model chosen by the data disseminators, we conclude that it is the preferred microdata synthesis approach.  

Future extensions of this work will include multiple variables / responses (where we would extend FBS and FBP synthesizers to the modeling of a multivariate response together with the survey weight variable to privacy protect the joint distribution of these sensitive variables), two-stage sample surveys, and alternative formulations of DP for Bayesian methods; for example, censoring / transforming the loglikelihood to ensure global DP for all sample sizes \citep{SavitskyWilliamsHu2020ppm}.

As noted earlier, there are more advanced additive noise processes than that we used \citep[For example][]{li2010optimizing, li2014DAWA}, which may lead to some efficiencies for the additive noise method, but they are complicated to implement and require specialized optimization software unavailable to the authors of this paper.

\section*{Acknowledgement}
This research was supported, in part, by the National Science Foundation (NSF), National Center for Science and Engineering Statistics (NCSES) by the Oak Ridge Institute for Science and Education (ORISE) for the Department of Energy (DOE). ORISE is managed by Oak Ridge Associated Universities (ORAU) under DOE contract number DE-SC0014664. 

The authors also wish to thank Phillip Leclerc, a Mathematical Statistician at the U.S. Census Bureau, for his guidance and advice on our implementation of the added noise Laplace Mechanism.

All opinions expressed in this paper are the authors' and do not necessarily reflect the policies and views of NSF, BLS, DOE, ORAU, or ORISE.


\vspace{10mm}
\appendix
\noindent {\huge{\bf{Appendix}}}

\section{Review of combining rules for partial synthesis}
\label{sec:appendix:combiningrules}

For estimand $Q$, let $q^{(\ell)}$ be the point estimator $q$ of $Q$,  and $u^{(\ell)}$ be variance of $q$ in $\ell$th synthetic dataset $(\bm y_n^{*, (\ell)}, \bm X_n^{(\ell)}, \bm w_n^{*, (\ell)})$. The analyst can use $\bar{q}_m = \sum_{\ell = 1}^{m}q^{(\ell)}/m$ to estimate $Q$, and $T = b_m / m + \bar{u}_m$ to estimate the variance of $\bar{q}_m$, where $b_m = \sum_{\ell = 1}^{m} (q^{(\ell)} - \bar{q}_m)^2 / (m - 1)$ and $\bar{u}_m = \sum_{\ell = 1}^{m} u^{(\ell)} / m$, where $u^{(\ell)}$ denotes the within database $\ell$ variance for $\ell = 1,\ldots,m$. Inferences can be based on $t$ distributions with degrees of freedom $v = (m - 1) (1 + \bar{u}_m / (b_m / m))^2$ when the sample size is large.

\newpage

\title{\large \bf Supplementary Material for Private Tabular Survey Data Products through Synthetic Microdata Generation}

\abstract{We provide detailed review of the Laplace Mechanism, specification of the prior distributions used in the two risk-weighted Bayesian models presented in the main text, Sections 2.1 and 2.2. We also provide additional weight smoothing effects results in Section 4.2 from the main text. Lastly, results of comparing FBS and FBP under repeated sampling are presented, including coverage vs CV plots where we explore different $\epsilon$ values and detailed cell-by-cell Monte Carlo simulation results of $(m = 3, \epsilon = 10.8)$.  }

\section{Review of differentially private mechanisms}

In this section we describe some of the major classes of mechanisms for data release that can be shown to be differentially private (i.e. satisfy the property in Definition 1 in the main text). In particular, we describe additive noise, the Exponential Mechanism, and the Bayesian posterior mechanism and discuss connections between the three. Variations of these approaches will be compared in our analyses.

Perhaps the simplest way to protect a data release is to generate random noise and add it the original data value (record level or tabular). The key insight from \citet{Dwork:2006:CNS:2180286.2180305} is that we can calibrate the amount of noise (scale or variability) to meet specific target values of $\epsilon$. For smaller $\epsilon$, more noise (larger scale) is needed.
The ``right" amount of noise to add is based on the \emph{sensitivity} of the mechanism $\mathcal{M}$.

\begin{definition}\label{def:sensitivity}
Let $D, D' \in \mathbb{R}^{k \times n}$. Let $q$ define a (non-differentially private) data release (e.g. mean), $q(): \mathbb{R}^{k \times n} \rightarrow S$. Then define the L1- sensitivity of $q$ as
\[
\Delta_{q} = \sup_{D, D' \in \mathbb{R}^{k \times n}}\left\| q(D) -  q(D') \right\|_{1}
\]
\end{definition}
The sensitivity of the desired data release mechanism $q()$ tells us how much noise to add to get a differentially private version of the mechanism $\mathcal{M}()$. The most common example is the Laplace Mechanism, which adds noise from a Laplace distribution. The density of a Laplace distribution is the following:
\[
f_{\mathcal{LAP}}(x \mid \mu, b) = \frac{1}{b}\exp\left( - \frac{1}{b} \abs{x - \mu} \right)
\]
Given a deterministic data release $q(): \mathbb{R}^{k \times n} \rightarrow S$, define a Laplace Mechanism $\mathcal{M}_{\mathcal{L}}() = q() + \mathcal{LAP}(0, \Delta_{q}/\epsilon)$. Then $\mathcal{M}_{\mathcal{L}}$ is an $\epsilon$-differentially private release mechanism \citep{Dwork:2006:CNS:2180286.2180305}.

\citet{Dwork:2006:CNS:2180286.2180305} also considers metrics beyond the absolute difference measure (L1) to measure sensitivity and how to create a mechanism that is differentially private. Subsequent works clarified this formulation and named it the Exponential Mechanism and showed how the Laplace Mechanism is a special case \citep{McSherryTalwar2007, WassermanZhou2010}.

Switching notation, slightly, let $\theta$ be the desired output (previously $s$) which could be a synthetic value, a model parameter, or a tabular summary statistic.
The Exponential Mechanism inputs a non-private mechanism for $\theta$ and generates $\theta$ in such a way that induces a DP guarantee on the overall mechanism.
\begin{definition}
The Exponential Mechanism $\mathcal{M}_{\mathcal{E}}$ uses a utility function $u(D, \theta)$ to generate values of $\theta$ from a distribution proportional to,
\begin{equation}\label{ref:EMdef}
\mathcal{M}_{\mathcal{E}}(\theta \mid D) \sim \exp \left(\frac{\epsilon \, u(D, \theta)}{2 \Delta_{u}} \right)\xi\left(\theta\mid \gamma\right),
\end{equation}
where, $\Delta_{u} = \mathop{\sup}_{D, D' \in \mathbb{R}^{k \times n}: \delta(D, D') = 1} \, \, \mathop{\sup}_{\theta \in \Theta} \,\, \abs{u(D, \theta) - u(D', \theta)}$ is the sensitivity, defined globally over $D \in \mathbb{R}^{k \times n}$, $\delta(D, D') = \# \{i: D{i} \neq D'{i} \}$ which is the Hamming distance between $D, D' \in \mathbb{R}^{k \times n}$.  $\xi\left(\theta\mid \gamma\right)$ is a proper probability measure.
\end{definition}
Each single draw of $\theta$ from the Exponential Mechanism $\mathcal{M}_{\mathcal{E}}(\theta|D)$ satisfies $\epsilon$-DP. See \citet{McSherryTalwar2007} or \cite{WassermanZhou2010}.

The main benefit of the Exponential Mechanism over (symmetric) additive noise is that the general utility function $u(D, \theta)$ can be constructed in such a way that restrictions to the range of output (for example enforcing positive counts) are possible and the input data can be less restrictive (for example variables without natural bounds, such as revenue). In other words, the perturbation is not symmetric by default and can be applied to more variable types besides categories and counts.
A major challenge when using the Exponential Mechanism is actually sampling from the implied distribution in (\ref{ref:EMdef}). For stability, a base (or prior) probability measure is often used to guarantee that implied distribution is a proper probability measure ($\xi\left(\theta\mid \gamma\right)$ integrates to 1 instead of $\infty$). Still, generating a $\theta$ using an arbitrary utility function $u()$ is a significant challenge \citep{WassermanZhou2010, SnokeSlavkovic2018PSD}. 

\citet{pmlr-v37-wangg15} show the connection between the Exponential Mechanism and sampling from the posterior distribution of a probability model by setting the utility $u$ as the log-likelihood. The main benefit of using Bayesian methods is the extensive research into computational methods to generate samples, something that presents a significant challenge for the general Exponential Mechanism.
\cite{Dimitrakakis:2017:DPB:3122009.3122020} provide alternative extensions and proofs for the differential privacy property of samples from the posterior distribution. They also assume the log-likelihood is bounded and suggest truncating the support for $\theta$ to achieve this.
\citet{SavitskyWilliamsHu2020ppm} extend these results to incorporate individual level adjustments, where the weights $\alpha_i \propto \frac{1}{\hat{\Delta}_i}$ are related to record-specific sensitivity estimates $\hat{\Delta}_i$.
\begin{equation}
\xi^{\bm{\alpha}}\left(\theta \mid \mathbf{x},\gamma\right) \propto \left[\mathop{\prod}_{i=1}^{n}p\left(x_{i} \mid \theta\right)^{\alpha_i }\right]\xi\left(\theta\mid \gamma\right).
\label{eq:weighted}
\end{equation}

\section{Prior specification of FBS in Section 2.1}

We specify an independent and identically-distributed multivariate Gaussian prior for the coefficient locations, $(\boldsymbol{\beta}_y)$ and $(\boldsymbol{\beta}_w)$, in Equations (\ref{eq:betay_prior}) and (\ref{eq:betaw_prior}):
\begin{eqnarray}
    \bm \beta_y &\overset{iid}{\sim}& \textrm{MVN}_K(\bm 0, \textrm{diag}(\bm \sigma_{\beta_y}) \times \Omega_{\beta} \times \textrm{diag}(\bm \sigma_{\beta_y})), \label{eq:betay_prior}\\
    \bm \beta_w &\overset{iid}{\sim}& \textrm{MVN}_K(\bm 0, \textrm{diag}(\bm \sigma_{\beta_w}) \times \Omega_{\beta} \times \textrm{diag}(\bm \sigma_{\beta_w})), \label{eq:betaw_prior}
\end{eqnarray} 
where $K$ is the dimension of the predictors $\bm x_i$, and $\Sigma_{\beta}$ is a $K \times K$ correlation matrix. We give $\Sigma_{\beta}$ a Cholesky factor of 6 and each component of $\bm \sigma_{\beta_y}$ and $\bm \sigma_{\beta_w}$ receives a student-t prior with 3 degrees of freedom and scale 10.

For covariance parameter $\Sigma$, we specify an LKJ prior as in Equation (\ref{eq:Sigma_prior}):
\begin{equation}
    \Sigma = \textrm{diag}(\bm \sigma_{\Sigma}) \times \Omega_{\Sigma} \times \textrm{diag}(\bm \sigma_{\Sigma}), \label{eq:Sigma_prior}
\end{equation}
where $d = 2$ is the dimension of the response vector $[y_i, w_i]$ and $\Omega_{\Sigma}$ is a $d \times d$ correlation matrix. We give $\Omega_{\Sigma}$ a Cholesky factor of 6 and each component of $\bm \sigma_{\Sigma}$ receives a student-t prior with 3 degrees of freedom and scale 10.

We refer interested readers to \citet{Rstan} for the usage and specification of LKJ priors.

\section{Prior specification of FBP in Section 2.2}

 We specify independent multivariate normal priors for $\bm \beta$ and $(\kappa_y, \bm \kappa_x)$, in Equations (\ref{eq:betaFBP_prior}) and (\ref{eq:kappa_prior}):
 \begin{eqnarray}
     \bm \beta &\sim& \textrm{MVN}(\bm 0, 100 \bm I), \label{eq:betaFBP_prior} \\
     (\kappa_y, \bm \kappa_x) &\sim& \textrm{MVN}(\bm 0, 100 \bm I), \label{eq:kappa_prior}
 \end{eqnarray}
 where $\bm I$ is the identity matrix. 

For $\sigma_y$ and $\sigma_x$, we specify a half Cauchy prior, in Equation (\ref{eq:sigma_prior}):
\begin{equation}
    \sigma_y, \sigma_x \overset{iid}{\sim} \textrm{Cauchy}^{+}(0, 1), \label{eq:sigma_prior}
\end{equation}
which is restricted to the positive real line.

\section{Additional results for weight smoothing effects in Section 4.2}

\begin{figure}[H]
  \centering
    \includegraphics[width=0.6\textwidth]{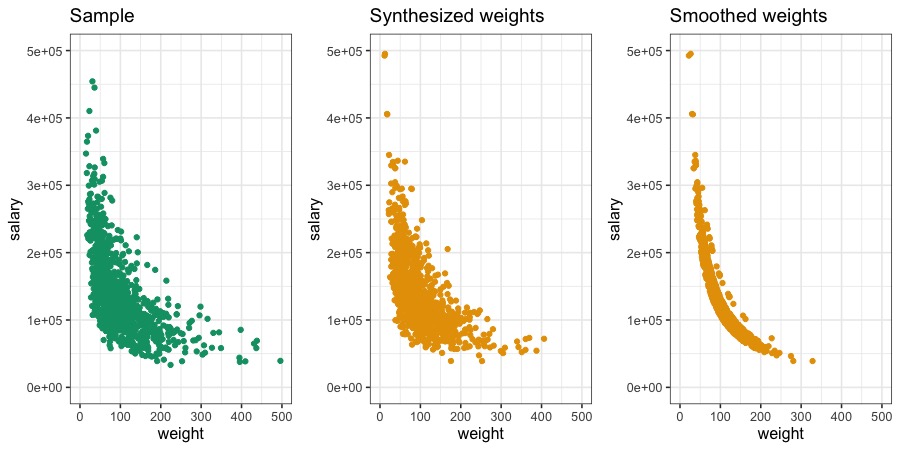}
      \caption{Salary versus sampling weights in the original, confidential sample (left), in one synthetic dataset with synthesized weights (middle), and in one synthetic dataset with smoothed weights (right) for FBS.}
      \label{fig:sim:FBS}
\end{figure}

\begin{figure}[H]
  \centering
    \includegraphics[width=0.6\textwidth]{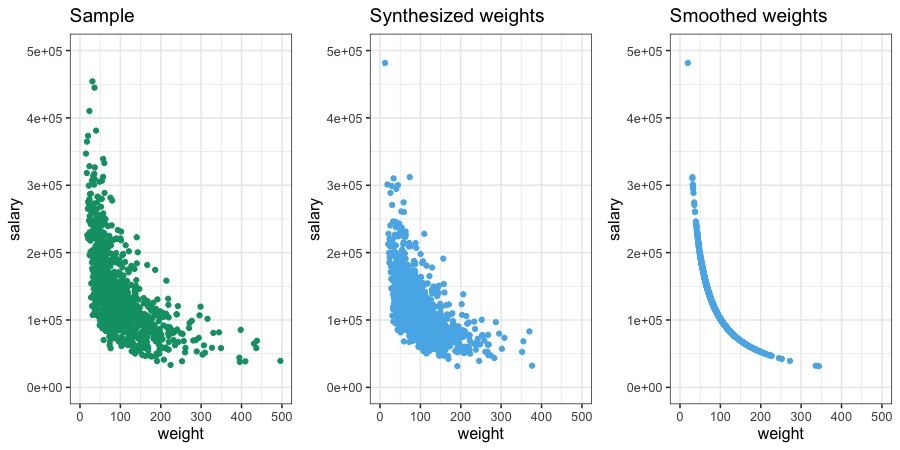}
      \caption{Salary versus sampling weights in the original, confidential sample (left), in one synthetic dataset with synthesized weights (middle), and in one synthetic dataset with smoothed weights (right) for FBP.}
      \label{fig:sim:FBP}
\end{figure}

\section{Comparison of FBS and FBP through repeated sampling}

To compare the performances of FBS and FBP and under repeated sampling, we conduct a Monte Carlo simulation study by generating $R = 100$ samples of size $n = 1000$ from our simulated population containing $N = 100,000$ units, described in Section 4.1 in the main text. Our goal here is to compare the competing two methods, FBS and FBP, acknowledging that for each of them the nominal coverage is not likely to be achieved because (1) we have a moderate fixed sample size and more importantly (2) each method induces a mis-specification and perturbation of the original generating distribution to enhance privacy protection. We do not attempt to demonstrate precise nominal coverage under an asymptotic frequentist claim and assume correct model specification, which would require to use a series of larger sample sizes and a much larger number of replicates to control the Monte Carlo error to much higher precision, that is computationally infeasible in our setting. In other words, there may be a concern about Monte Carlo standard error from the relatively low number of iterations ($R = 100$), but \citep{LeonSavitsky2019EJS} estimated very similar models to FBP under a range of Monte Carlo iterations and found little-to-no difference in the comparison of results, which indicates that the magnitude of Monte Carlo standard error is small.

For each of the $r = 1, \cdots, R$ sample, we apply the FBS and FBP approaches, and scale them through $(c_1, c_2)$ to have equivalent $(\epsilon_{\bm y_n} = 2 \Delta_{\bm \alpha, (\bm y_n, \bm X_n, \bm w_n)} \times m)$, where we simulate $m = 3$ synthetic datasets from each approach. In fact, from now on we will use $\epsilon$ instead of $\epsilon_{\bm y_n}$ to indicate that we target each $\epsilon_{\bm y_n}$ to achieve the target global $\epsilon$, which mimics real world application of our pseudo posterior synthesizers. Through scaling and shifting, we are able to achieve the targeted overall Lipschitz bound $\Delta_{\bm \alpha, (\bm y_n, \bm X_n, \bm w_n)}$ in every simulated database, and therefore the $\epsilon$ guarantee is global.

\begin{figure}[H]
  \centering
    \includegraphics[width=0.7\textwidth]{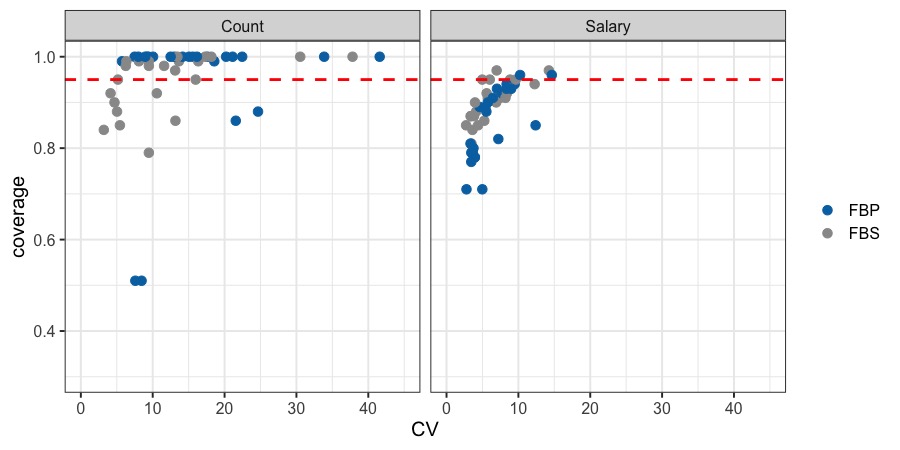}
      \caption{$\boldsymbol{(m = 3, \epsilon = 10.8):}$ Coverage vs CV of {\bf{counts}} (left) and {\bf{average salary values}} (right) of FBS and FBP, over $R = 100$ replicates. A red dashed line at coverage = 0.95 is included for reference. Each dot represents a coverage vs CV result for one of 24 cells. Results are based on $m = 3$ synthetic datasets by FBS and FBP, achieving $\epsilon = 10.8$ for both methods. On average, the confidence interval based on FBP is 1.38 times of that of FBS for {\bf{counts}}, and 0.97 times for {\bf{average salary values}}.}
      \label{fig:sim_MC_Count_Salary_0p3to1_m3_L1p8}
\end{figure}

For utility evaluation, we calculate the effective coverage rate of the nominal 95\% confidence interval of the population count and average salary by field and gender. The closer the coverage rate to 0.95, the better the performance. In addition, we calculate the average ratio of the coefficients of variation (CV), which are the standard errors scaled by the corresponding estimates, and compare between the two approaches. The CV provides information about the relative efficiency, and the lower the CV, the better the performance. The detailed cell-level results of $m = 3$ are available in Tables \ref{tab:covCI_count_m3} and \ref{tab:covCI_salary_m3}. Here, we use Figure \ref{fig:sim_MC_Count_Salary_0p3to1_m3_L1p8} to display coverage of the $95\%$ nominal interval on the x-axis as compared to CV on the y-axis for cell counts and average salary, respectively. Each point represents one of the 24 cells: $8 \times 2 = 16$ cells for a field and gender combination, and 8 cells for both genders for 8 fields.

For results of counts shown on the left of Figure \ref{fig:sim_MC_Count_Salary_0p3to1_m3_L1p8}, FBP has higher coverage than FBS in most of the cells. The two dots showing its serious undercoverage issue correspond to both genders and male alone of Field 7 (both coverage rates are 0.51). The higher coverage of FBP comes at the cost of wider average confidence intervals: on average, the length of the confidence intervals of FBP is 1.38 times of that of FBS (see Table \ref{tab:covCI_count_m3} for cell-by-cell comparison). These results indicate that FBP overcovers and produces overly long intervals, which is sometimes desirable by being conservative. By contrast, FBS only slightly undercovers, and achieves greater efficiency with smaller variances and shorter intervals. 



For results of average salary values shown on the right of Figure \ref{fig:sim_MC_Count_Salary_0p3to1_m3_L1p8}, FBS and FBP perform equally well in terms of coverage efficiency manifested by CV, and FBS performs slightly better in terms of coverage rates. It is worth noting that on average, the length of the confidence intervals of FBP is 0.95 times of that of FBS (see Table \ref{tab:covCI_salary_m3} for cell-by-cell comparison). The longer confidence intervals produced by FBS could partially explain its slightly better coverage performance. Nevertheless, as we have seen in the results of one replicate of simulated sample in Sections 4.3 and 4.4 in the main text, FBP performs better than FBS on average salary values since it incorporates correction for population bias of the salary outcome variable. Our Monte Carlo simulation result reinforces this finding.

In summary, FBS achieves greater efficiency for counts, whereas the two approaches perform overall equally well for average salary values.

\begin{table}[H]
\centering
\begin{tabular}{ l | c | c c | c c }
\hline
 & Pop. & FBS coverage &   avg CI & FBP coverage &  avg CI   \\ \hline
 Field 1 total & 27963     & 0.84 & 3560  & 0.99 & 6682   \\ 
 Male  & 14599   & 0.99 & 3705  & 1.00 & 4854 \\ 
 Female &13364  &  0.99 & 4298  & 1.00 & 5242 \\ \hline
 Field 2 total & 2734    & 0.86 & 1416  & 0.86 & 1853 \\ 
 Male  & 1951  & 0.95 & 1287 & 0.88 & 1492     \\ 
 Female & 783  & 1.00 & 1185   & 1.00 & 1037    \\ \hline
 Field 3 total & 4035  & 0.92 & 1709  & 1.00 & 2814    \\ 
 Male  & 2918 & 0.97 & 1576   & 1.00 & 2258  \\ 
 Female  & 1117 & 1.00 & 1446     & 1.00 & 1596   \\  \hline
 Field 4 total & 18109   & 0.90 & 3312  & 1.00 & 5280    \\ 
 Male  & 12381 & 0.98 & 3175 & 1.00 & 4320   \\ 
 Female  & 5728 & 0.99 & 3025  & 1.00 & 3360  \\  \hline
 Field 5 total & 11197   & 0.85 & 2474  &  1.00 & 4876    \\ 
 Male  & 3861& 1.00 & 2081  & 1.00 & 2713   \\ 
 Female  & 7336 & 0.98 & 2781  & 1.00 & 4104 \\ \hline
 Field 6 total & 12842 & 0.88 & 2576  &  1.00 & 4915 \\ 
 Male  & 6433 & 0.99 & 2488   & 1.00 & 3393  \\ 
 Female  & 6409 & 0.98 & 2892    & 1.00 & 3702  \\ \hline
 Field 7 total & 18131   & 0.92 & 2955 &  0.51 & 4725 \\ 
 Male & 14358  & 0.95 & 3021  & 0.51 & 4108 \\ 
 Female  & 3773  & 1.00 & 2691 &  0.99 & 2516    \\ \hline
 Field 8 total & 4989  & 0.79 & 1934 & 1.00 & 2936  \\ 
 Male  & 2033 & 1.00 & 1557 &  1.00 & 1760 \\
 Female  & 2956 & 0.99 & 1908 & 1.00 & 2297 \\ \hline
\end{tabular}
\caption{$\boldsymbol{(m = 3, \epsilon = 10.8):}$ Coverage of counts and average length of confidence intervals of FBS and FBP under 100 repeated samples. On average, the confidence interval based on FBP is 1.38 times of that of FBS.}
\label{tab:covCI_count_m3}
\end{table}

\begin{table}[H]
\centering
\begin{tabular}{ l | c | c c | c c }
\hline
 & Pop. & FBS coverage &   avg CI & FBP coverage &  avg CI   \\ \hline
 Field 1 total & 110659  & 0.85 & 13657  & 0.71 & 13422  \\ 
 Male  & 121125 & 0.87 & 18337   & 0.77 & 17772 \\ 
 Female   & 99226  & 0.90 & 17941 & 0.78 & 16461  \\ \hline
 Field 2 total & 140742 & 0.95 & 58335    & 0.94 & 64240    \\ 
 Male  & 146916 &  0.95 & 66324   & 0.96 & 71592   \\ 
 Female  & 125358  & 0.97 & 78486  & 0.96 & 78408 \\  \hline
 Field 3 total & 114668  & 0.91 & 41026   & 0.92 & 41596  \\ 
 Male  & 118338  & 0.91 & 47514 & 0.93 & 46015 \\ 
 Female  & 105081 & 0.94 & 56496  & 0.85 & 53166   \\ \hline
 Field 4 total & 115647  & 0.81 & 19026   & 0.81 & 16953  \\ 
 Male  & 122601 & 0.87 & 22022   & 0.80 & 20016   \\ 
 Female  & 100618  & 0.91 & 25821   & 0.90 & 23786 \\ \hline
 Field 5 total & 105808 & 0.85 & 21648    & 0.89 & 22418  \\ 
 Male  & 120531  & 0.95 & 31905  & 0.91 & 33535   \\ 
 Female  & 98060 &   0.86 & 23600 & 0.88 & 24292  \\ \hline
 Field 6 total & 110159   & 0.88 & 21325   & 0.78 & 18347  \\ 
 Male  & 122676  & 0.95 & 27187  & 0.71 & 25241    \\ 
 Female  & 97595  & 0.92 & 25165    & 0.89 & 21989   \\ \hline
 Field 7 total & 132249  &  0.87 & 21163   & 0.79 & 20262   \\ 
 Male  & 136370  & 0.84 & 22964  & 0.79 & 22021  \\ 
 Female  & 116566  & 0.97 & 34669  & 0.82 & 34129  \\ \hline
 Field 8 total & 119005  & 0.90 & 40658 & 0.93 & 39461 \\  
 Male &137349  & 0.93 & 57063  & 0.93 & 54326 \\
 Female  & 106388  & 0.92 & 43357  & 0.94 & 40191  \\  \hline
\end{tabular}
\caption{$\boldsymbol{(m = 3, \epsilon = 10.8):}$ Coverage of average salary values and average length of confidence intervals of FBS and FBP under 100 repeated samples. On average, the confidence interval based on FBP is 0.97 times of that of FBS.}
\label{tab:covCI_salary_m3}
\end{table}

\begin{figure}[H]
  \centering
    \includegraphics[width=0.7\textwidth]{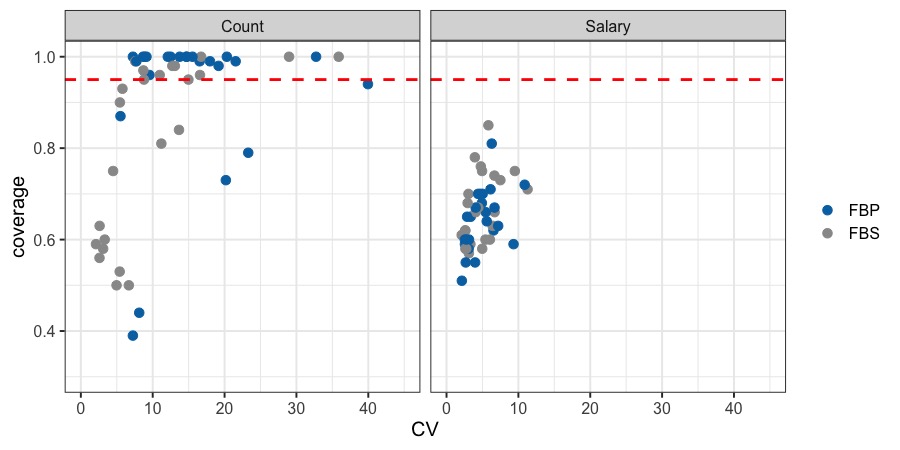}
      \caption{$\boldsymbol{(m = 1, \epsilon = 3.6):}$ Coverage vs CV of {\bf{counts}} (left) and {\bf{average salary values}} (right) of FBS and FBP, over $R = 100$ replicates. A red dashed line at coverage = 0.95 is included for reference. Each dot represents a coverage vs CV result for one of 24 cells. Results are based on $m = 1$ synthetic dataset by FBS and FBP, achieving $\epsilon = 3.6$ for both methods. On average, the confidence interval based on FBP is 1.81 times of that of FBS for {\bf{counts}}, and 0.93 times for {\bf{average salary values}}.}
      \label{fig:sim_MC_Count_Salary_0p3to1_m1_L1p8}
\end{figure}

\begin{figure}[H]
  \centering
    \includegraphics[width=0.7\textwidth]{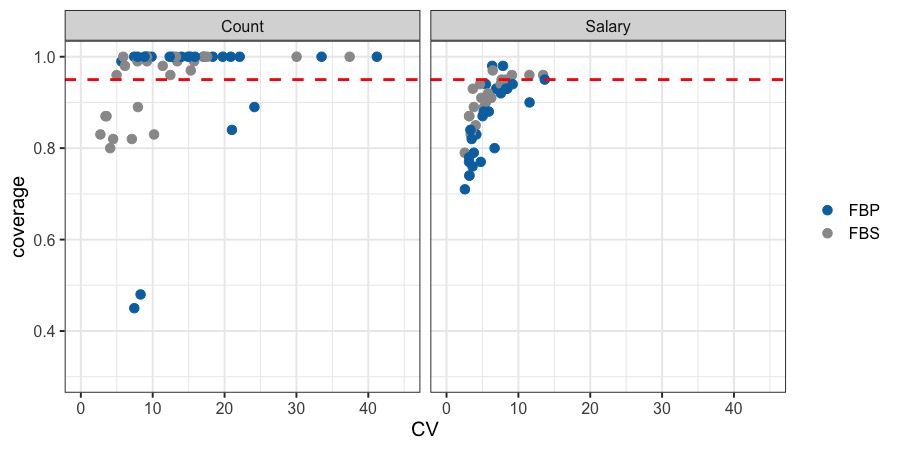}
      \caption{$\boldsymbol{(m = 5, \epsilon = 18):}$ Coverage vs CV of {\bf{counts}} (left) and {\bf{average salary values}} (right) of FBS and FBP, over $R = 100$ replicates. A red dashed line at coverage = 0.95 is included for reference. Each dot represents a coverage vs CV result for one of 24 cells. Results are based on $m = 5$ synthetic dataset by FBS and FBP, achieving $\epsilon = 18$ for both methods. On average, the confidence interval based on FBP is 1.54 times of that of FBS for {\bf{counts}}, and 0.95 times for {\bf{average salary values}}.}
      \label{fig:sim_MC_Count_Salary_0p3to1_m5_L1p8}
\end{figure}

\begin{figure}[t!]
  \centering
    \includegraphics[width=0.7\textwidth]{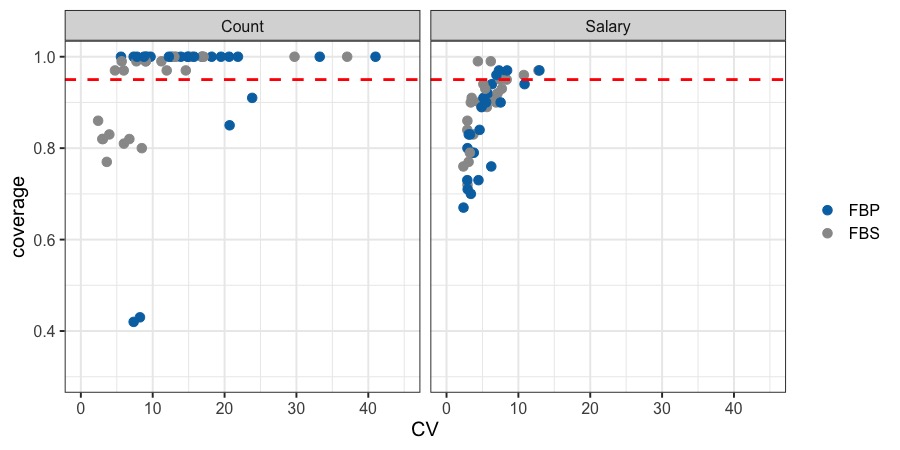}
      \caption{$\boldsymbol{(m = 10, \epsilon = 36):}$ Coverage vs CV of {\bf{counts}} (left) and {\bf{the average salary}} (right) of FBS and FBP, over $R = 100$ replicates. A red dashed line at coverage = 0.95 is included for reference. Each dot represents a coverage vs CV result for one of 24 cells. Results are based on $m = 10$ synthetic dataset by FBS and FBP, achieving $\epsilon = 36$ for both methods. On average, the confidence interval based on FBP is 1.65 times of that of FBS for {\bf{counts}}, and 0.95 times for {\bf{the average salary}}.}
      \label{fig:sim_MC_Count_Salary_0p3to1_m10_L1p8}
\end{figure}

We also evaluate the effects of $m$, the number of simulated synthetic datasets, on the utility-risk trade-off under repeated sampling. Figure \ref{fig:sim_MC_Count_Salary_0p3to1_m1_L1p8} shows the coverage vs CV results between FBS and FBP for $(m = 1, \epsilon = 3.6)$, and Figure \ref{fig:sim_MC_Count_Salary_0p3to1_m5_L1p8} shows results for $(m = 5, \epsilon = 18)$. Simulating only $m = 1$ synthetic dataset creates more serious undercoverage issues for both models for counts and average salary values, suggesting too much a utility reduction compromise at a lower privacy budget. When simulating $m = 5$ synthetic datasets at the price of a higher privacy budget, there is slight improvement of coverage performance of both models for average salary values, although FBP still has an overcoverage issue with most of the cells and the serious undercoverage issue for the two cells related to Field 7. These results suggest simulating $m = 3$ synthetic datasets achieves a good balance of utility-risk trade-off for both models, with reasonably high utility at a reasonable privacy budget of $\epsilon = 10.8$. Further increasing $m$ for $(m = 10, \epsilon = 36)$ produces similar utility results compared to that of $(m = 5, \epsilon = 18)$, as Figures \ref{fig:sim_MC_Count_Salary_0p3to1_m5_L1p8} and \ref{fig:sim_MC_Count_Salary_0p3to1_m10_L1p8} illustrate. 

\begin{figure}[H]
  \centering
    \includegraphics[width=0.7\textwidth]{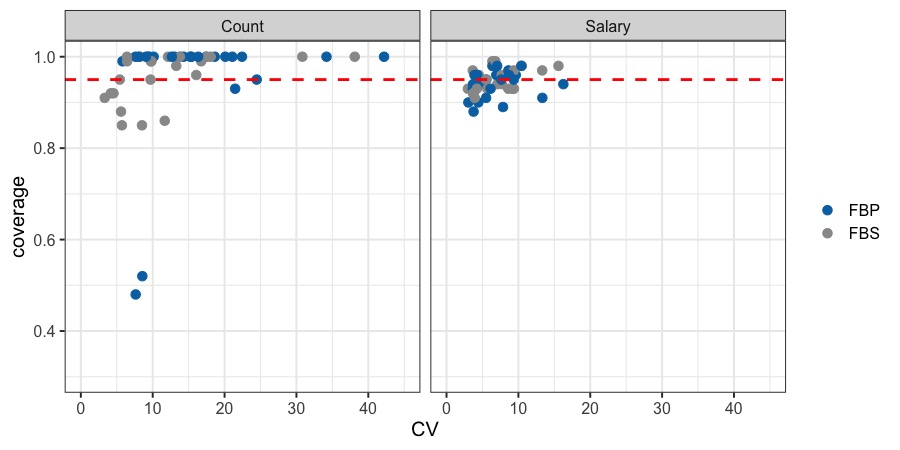}
      \caption{$\boldsymbol{(m = 3, \epsilon = 20.4):}$ Coverage vs CV of {\bf{counts}} (left) and {\bf{the average salary}} (right) of FBS and FBP, over $R = 100$ replicates. A red dashed line at coverage = 0.95 is included for reference. Each dot represents a coverage vs CV result for one of 24 cells. Results are based on $m = 3$ synthetic datasets by FBS and FBP, achieving $\epsilon = 20.4$ for both methods. On average, the confidence interval based on FBP is 1.39 times of that of FBS for {\bf{counts}}, and 0.97 times for {\bf{the average salary}}.}
      \label{fig:sim_MC_Count_Salary_0p3to1_m3_L3p4}
\end{figure}

\begin{figure}[H]
  \centering
    \includegraphics[width=0.7\textwidth]{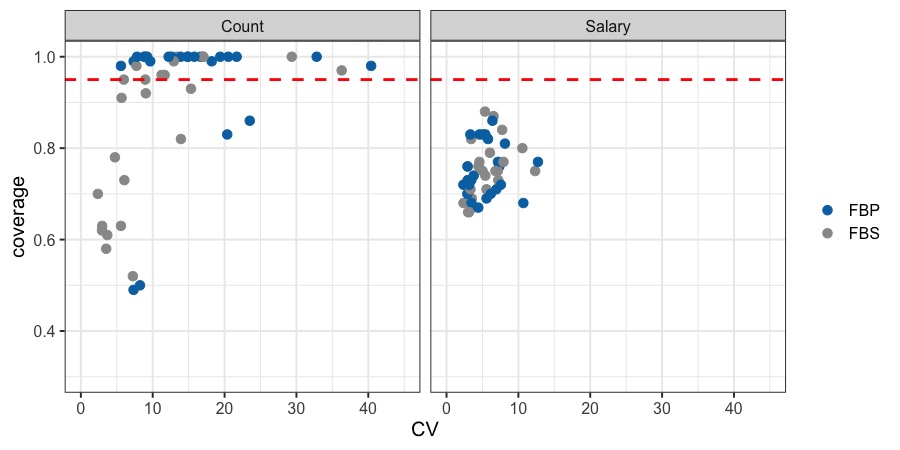}
      \caption{$\boldsymbol{(m = 1, \epsilon = 6.8):}$ Coverage vs CV of {\bf{counts}} (left) and {\bf{the average salary}} (right) of FBS and FBP, over $R = 100$ replicates. A red dashed line at coverage = 0.95 is included for reference. Each dot represents a coverage vs CV result for one of 24 cells. Results are based on $m = 1$ synthetic dataset by FBS and FBP, achieving $\epsilon = 6.8$ for both methods. On average, the confidence interval based on FBP is 1.71 times of that of FBS for {\bf{counts}}, and 0.97 times for {\bf{the average salary}}.}
      \label{fig:sim_MC_Count_Salary_0p3to1_m1_L3p4}
\end{figure}

\begin{figure}[H]
  \centering
    \includegraphics[width=0.7\textwidth]{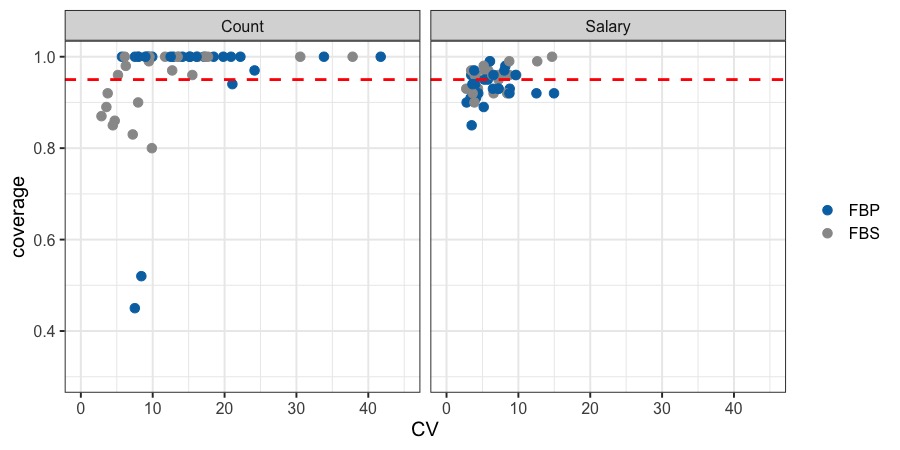}
      \caption{$\boldsymbol{(m = 5, \epsilon = 34):}$ Coverage vs CV of {\bf{counts}} (left) and {\bf{the average salary}} (right) of FBS and FBP, over $R = 100$ replicates. A red dashed line at coverage = 0.95 is included for reference. Each dot represents a coverage vs CV result for one of 24 cells. Results are based on $m = 5$ synthetic dataset by FBS and FBP, achieving $\epsilon = 34$ for both methods. On average, the confidence interval based on FBP is 1.52 times of that of FBS for {\bf{counts}}, and 0.97 times for {\bf{the average salary}}.}
      \label{fig:sim_MC_Count_Salary_0p3to1_m5_L3p4}
\end{figure}

\begin{figure}[H]
  \centering
    \includegraphics[width=0.7\textwidth]{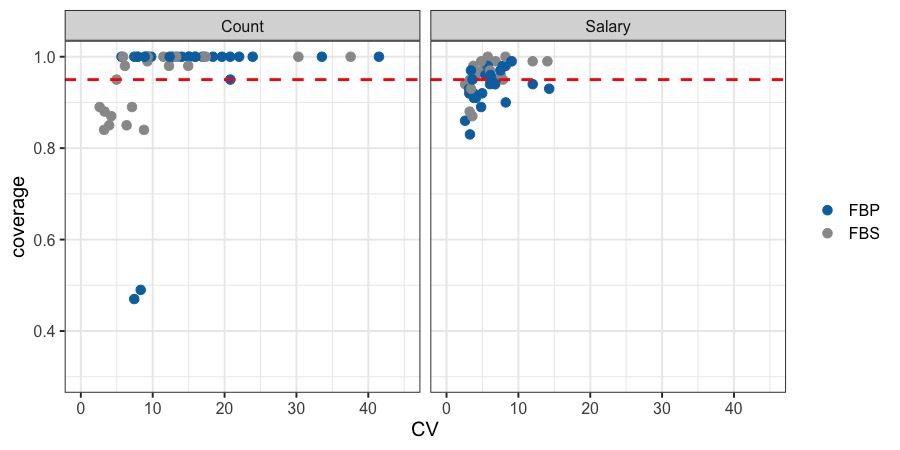}
      \caption{$\boldsymbol{(m = 10, \epsilon = 68):}$ Coverage vs CV of {\bf{counts}} (left) and {\bf{the average salary}} (right) of FBS and FBP, over $R = 100$ replicates. A red dashed line at coverage = 0.95 is included for reference. Each dot represents a coverage vs CV result for one of 24 cells. Results are based on $m = 10$ synthetic dataset by FBS and FBP, achieving $\epsilon = 68$ for both methods. On average, the confidence interval based on FBP is 1.60 times of that of FBS for {\bf{counts}}, and 0.96 times for {\bf{the average salary}}.}
      \label{fig:sim_MC_Count_Salary_0p3to1_m10_L3p4}
\end{figure}

Finally we present the results for $m = 3$ with an overall Lipschitz bound $\Delta_{\bm \alpha, (\bm y_n, \bm X_n, \bm w_n)} = 3.4$ in Figure \ref{fig:sim_MC_Count_Salary_0p3to1_m3_L3p4}. This is the case of $(m = 3, \epsilon = 20.4)$. Compared to $(m = 3, \epsilon = 10.8)$ in Figure \ref{fig:sim_MC_Count_Salary_0p3to1_m3_L1p8}, we observe notable improvement of both FBS and FBP on average salary values. Data disseminators could decide to release the results from $(m = 3, \epsilon = 20.4)$ given its superior utility performance, if they are comfortable with an increased privacy budget. Results of other $m$ values for $\Delta_{\bm \alpha, (\bm y_n, \bm X_n, \bm w_n)} = 3.4$ are available in Figures \ref{fig:sim_MC_Count_Salary_0p3to1_m1_L3p4} to \ref{fig:sim_MC_Count_Salary_0p3to1_m10_L3p4}.

\bibliographystyle{natbib}
\bibliography{DPbib}

\end{document}